\documentclass[onecolumn]{aa}
\usepackage{amsmath}
\usepackage{amssymb}
\usepackage{graphicx}
\usepackage{wasysym}
\usepackage{natbib}
\usepackage{color}

\newcommand{\pdb}[2]{\displaystyle \frac{\partial #1}{\partial #2}}

\newcommand{\dd}[1]{\mathrm{d}^{#1}}
\newcommand{\grad}[1]{\vf{\nabla}#1}
\newcommand{\vf}[1]{{\bf{#1}}}
\newcommand{\vfg}[1]{{\boldsymbol{#1}}}
\newcommand{\ten}[1]{\boldsymbol{\mathsf{#1}}}
\newcommand{\im}{\mathrm{i}}
\newcommand{\omt}{\widetilde{\omega}}
\newcommand{\Btheta}{B_{\vartheta}}

\newcommand{\vz}{v_{z}}
\newcommand{\operator}[1]{\mathcal{#1}}
\newcommand{\ppsi}{\partial_{\psi}}
\newcommand{\ptheta}{\partial_{\vartheta}}
\newcommand{\D}{\operator{D}}
\newcommand{\Dd}{\operator{D}^{\dagger}}
\newcommand{\dD}{^\dagger\operator{D}}
\newcommand{\F}{\operator{F}}
\newcommand{\G}{\operator{G}}
\newcommand{\kp}{\kappa_{\mathrm{p}}}
\newcommand{\kt}{\kappa_{\mathrm{t}}}
\newcommand{\kg}{\kappa_{\mathrm{g}}}
\newcommand{\matel}[1]{\mathrm{#1}}

\newcommand{\e}{\mathrm{e}}
\newcommand{\gp}{\gamma p}
\newcommand{\gpz}{\gamma p_{0}}
\newcommand{\gpB}{\gamma p + B^{2}}
\newcommand{\gpBz}{\gamma p_{0} + B_{0}^{2}}
\newcommand{\Nmpol}{N_{\mathrm{m,pol}}^{2}}
\newcommand{\orNmpol}{\overline{\rho N_{\mathrm{m,pol}}^{2}}}
\newcommand{\mcO}{\mathcal{O}}

\begin{document}

\title{Toward detailed prominence seismology}
\subtitle{II. Charting the continuous magnetohydrodynamic spectrum}
\author{J.W.S. Blokland\inst{1} \and R. Keppens\inst{2}}
\institute{FOM Institute for Plasma Physics Rijnhuizen, 
           Association EURATOM-FOM, 
	   P.O. Box 1207, 3430 BE Nieuwegein, The Netherlands
           \and
	   Centre for Plasma Astrophysics, Mathematics Department, K.U. Leuven, Celestijnenlaan 200B, 3001 Heverlee, Belgium}
\offprints{J.W.S. Blokland \email{J-W.S.Blokland@Rijnhuizen.nl}}
\date{Received / Accepted}
\abstract{
          Starting from accurate magnetohydrodynamic flux rope equilibria containing prominence condensations, we initiate a systematic survey of their linear eigenoscillations. 
          This paves the way for more detailed prominence seismology, which thus far has made dramatic simplifications about the prevailing magnetic field topologies.
         }
	 {
          To quantify the full spectrum of linear MHD eigenmodes, we require knowledge of all flux-surface localized modes, charting out the continuous parts of the MHD spectrum. 
          We combine analytical and numerical findings for the continuous spectrum for realistic prominence configurations, where a cool prominence is embedded in a hotter cavity, 
          or where the flux rope contains multiple condensations supported against gravity.
	 }
	 {
          The equations governing all eigenmodes for translationally symmetric, gravitating equilibria containing an axial shear flow, are analyzed, along with their flux-surface 
          localized limit. The analysis is valid for general 2.5D equilibria, where either density, entropy, or temperature vary from one flux surface to another. We 
          analyze the intricate mode couplings caused by the poloidal variation in the flux rope equilibria, by performing a small gravity parameter expansion. We contrast the 
          analytical results with continuous spectra obtained numerically.
	 }
	 {
          For equilibria where the density is a flux function, we show that continuum modes can be overstable, and we present the stability criterion for these convective continuum 
          instabilities. Furthermore, for all equilibria, a four-mode coupling scheme between an Alfv\'enic mode of poloidal mode number $m$ and three neighboring $(m-1,m,m+1)$ slow 
          modes is identified, occurring in the vicinity of rational flux surfaces. For realistically structured prominence equilibria, this coupling is shown to play an important 
          role, from weak to stronger gravity parameter $g$ values. The analytic predictions for small $g$ are compared with numerical spectra, and progressive deviations for 
          larger $g$ are identified. 
	 }
	 {
          The unstable continuum modes could be relevant for short-lived prominence configurations. The gaps created by poloidal mode coupling in the continuous spectrum need further 
          analysis, as they form preferred frequency ranges for global eigenoscillations. 
	 }
\keywords{solar physics, solar prominences -- Instabilities -- Magnetohydrodynamics (MHD) -- Plasmas}
\titlerunning{Charting the continuous MHD spectrum}
\authorrunning{J. W. S. Blokland and R. Keppens}
\maketitle

\section{Prominence seismology from MHD modeling}
Motivated by the recent interest and progress made in diagnosing oscillations in solar prominences, we have started an effort~\citep[][hereafter Paper I]{Blokland_2011A} 
to close the gap between observed prominence properties, and the models used to quantify their oscillations. From observations, increasingly detailed knowledge of 
thermodynamical properties in filaments is available. This relies on inverting spectroscopic data, where departures from local thermodynamic equilibrium in the 
radiative transfer through one- and two-dimensional prominence models are nowadays routinely incorporated~\citep{Labrosse10}. At the same time, high resolution 
observations~\citep{Linetal05}, computations of three-dimensional (3D) magnetohydrostatic equilibrium configurations without gravity~\citep{Aulanier02}, or modern magnetic 
field extrapolations from photospheric magnetograms~\citep{DeRosa09}, together with the application of spectro-polarimetric techniques~\citep{Lin98,bueno02}, have brought 
insight into the complex three-dimensional topology of the surrounding magnetic field. Simultaneously, increased spatio-temporal resolution has quantified a host of 
oscillation modes observed in prominences, with velocity amplitudes of a few km s$^{-1}$. Both short- and long-period signals, from minutes to hours, can be detected, 
and regions that oscillate coherently with similar periods allow us to quantify damping rates for seemingly global eigenoscillations~\citep{Terradas02}.  

Theoretical studies in prominence seismology have meanwhile identified the most likely candidates explaining these motions, in terms of fundamental sausage, kink, and other 
mode types present in the MHD spectrum of magnetized slabs or cylinders. To explain the observed damping of the eigenmodes, one can on the one hand study wave 
damping as a result of non-adiabatic effects, such as anisotropic, thermal conduction, and radiative losses. For cylindrically structured coronal loops and prominences, 
with prevailing helical fields, early work by \cite{Keppens93} demonstrated how these non-adiabatic effects can give rise to damped and/or overstable global, discrete 
non-adiabatic Alfv\'en waves. \cite{Soler08} revisited these effects for eigenmodes of more simplified, uniform cylinder models in homogeneous magnetic 
fields. By using uniform internal/external conditions, the possibility of wave resonant behavior was excluded artificially. This resonant behavior, where global waves 
have eigenfrequencies in the range of the MHD continuous spectrum, provides an alternative means of explaining the observed damping rates. The theory for resonant MHD 
waves is reviewed in many papers, e.g. \cite{Goossens10}, and references therein. In other reviews of prominence oscillations~\citep[e.g.][]{Mackay10,Arregui11}, many 
of the accumulated theoretical findings are discussed, and a striking observation is that all these efforts dramatically simplify the underlying magnetic topology, in 
order to allow analytic progress. Non-uniformity, and the associated possibility of resonant absorption, is at most introduced by varying the density 
profiles~\citep[e.g.][]{Soler10}, sticking to uniform field topologies. At the same time, efforts to account for the multi-stranded nature of actual solar filaments 
have also been initiated, in similarly simple magnetic fields~\citep{Luna10}.

To pave the way for more detailed prominence seismology, true MHD spectroscopy~\citep{Goedbloed_1993}, of the kind pioneered in fusion-related plasma physics, is 
called for. This relies on accurate computations of the underlying equilibria, with full account of the complex magnetic topology, and subsequent computation of its 
eigenspectrum. In our accompanying paper I, we started this effort by demonstrating our ability to compute accurate translationally symmetric flux rope equilibria 
containing prominence condensations, governed by the magnetohydrostationary equations balancing Lorentz forces, pressure gradients, and gravity. We here continue 
this effort, and use these equilibria, where we incorporate helical fields, axial shear flows, and varying plasma beta conditions consistent with observed filament 
properties. We here concentrate on quantifying all eigenmodes defining the continuous parts of the MHD eigenspectrum. This is a worthwhile effort in its own right, 
partly because the global, discrete modes are known to be organized about this continuous 
spectrum~\citep[see e.g. modern textbooks such as][]{Book_Goedbloed_2004,Book_Goedbloed_2010}, and because the possible damping of these global modes due to resonant 
absorption relies on knowledge of the frequency ranges that they occupy. 

Charting out the continuous MHD spectrum for realistic flux rope structures with prominences has never been attempted before, in sharp contrast to the vast body of 
knowledge on MHD eigenmodes in axisymmetric, tokamak configurations. Our effort considers flux ropes idealized to be invariant along the filament axis, with an outer 
cylindrical cross-section, for simplicity. Since the gravitational stratification will act to deform a pure cylindrical flux model to one with still nested, 
but downwards displaced flux surfaces, the resulting poloidal variation in the equilibrium quantities will induce coupling between the different branches of the 
continuous MHD spectrum. These branches are labeled by their axial, $k$, and azimuthal, $m$, mode numbers. When charting the mode frequencies from the filament axis 
to the flux rope edge, the spectrum can help us to determine the avoided crossings, which are akin to the poloidal mode couplings discussed for equilibria without gravity, 
such as the nested elliptic cylinders studied by~\cite{Chance77}, or configurations with non-circular cross-sections and/or density variations along field lines analyzed 
by~\cite{Poedts_1991}. We extend these latter studies to translationally invariant, gravitating equilibria with possibly shear flow $v_z(\psi)$, restricted to be 
a flux function. The actual equilibria are presented and discussed in paper I. The analysis of their continuous spectra presented here, benefits from the detailed 
knowledge of the spectrum of the more complex, axisymmetric topologies~\citep{Goedbloed_1975}. The case for self-gravitating, static axisymmetric magnetized tori was 
treated analytically in~\cite{Poedts_1985}. \cite{Goedbloed_2004A} analyzed of the continuous MHD spectrum for magnetized accretion tori, where external 
gravity, nested toroidal flux surfaces, and both poloidal and toroidal (Keplerian-like) rotation is included, and identified a potential for unstable localized modes 
called trans-slow Alfv\'en continuum modes. The approach of~\cite{Goedbloed_2004A}, which combined analytic findings for the equilibria, the poloidal mode 
couplings that were confronted with numerically computed equilibria, and eigenmodes is analogous to our efforts in this series. 

Our paper is then organized as follows. In Sect.~\ref{sec:spectral}, we present the equations governing all eigenmodes for these gravitating, 2.5D MHD equilibria. 
Since we allow for axial flow, the Frieman-Rotenberg formalism~\citep{Frieman_1960} is revisited, and using the straight field line coordinates (see paper I) 
describing our equilibria, we present them in a form where a field line projection can be made. This form can directly be compared with all other cases studied 
previously, such as the cylindrical limit, or the analogy with tokamak configurations that contain toroidal flow. In Section~\ref{sec:continuous}, we then derive the equations 
for the continuous spectrum, and find that prominence equilibria where density is a flux function can also be affected by convective continuum 
instabilities~\citep{Blokland_2007B,vanderHolst_2000A,vanderHolst_2000B}. The coupled system of differential equations can be converted to algebraic relations between 
poloidal Fourier harmonics, which clearly identifies how Alfv\'en and slow modes can couple. This coupling is then studied analytically in Section~\ref{sec:expand}, 
where the small gravity parameter expansion previously used (paper I) to analyze equilibrium effects due to gravity, re-enters to analyze the leading order coupling 
effects. For low frequencies, we then demonstrate that a gap can arise in the Alfv\'en continuum branches, from a four-mode coupling scheme. In 
Section~\ref{sec:phoenix}, we briefly summarize the numerical approach used in the remaining sections. There, the equilibria presented in paper I are all diagnosed 
spectroscopically, by charting out their continuous spectrum. The analytic findings are used to check and analyze in detail the derived eigenfrequencies. This is 
done for cool prominences embedded in hot cavities in section~\ref{sec:cool}, and for complex, double layered prominences without and with axial flows in 
section~\ref{sec:double}. Conclusions and an outlook to future work is given in section~\ref{sec:conclusions}.
 
\section{Spectral formulation \label{sec:spectral}}
In what follows, we analyze the equations governing linear, ideal MHD perturbations about a stationary equilibrium. The equations for the equilibrium have been 
presented and solved numerically in paper I. The (time-independent) equilibrium quantities are denoted by the magnetic field vector $\vf{B}$, the pressure $p$, 
the density $\rho$, and the external gravitational potential $\Phi$, which for the prominence application at hand are all functions of the poloidal coordinates $(x,y)$. 
The background equilibrium may contain a purely longitudinal shear flow ${\bf{v}}=v_z(\psi){\bf{e}}_z$, where the axial flow is constant on each flux surface. Each 
flux label $\psi$ identifies a single flux surface, which form nested surfaces defined by the magnetic field topology within the prescribed outer flux rope shape. 
We first recapitulate the governing equations, and then present a representation exploiting a projection involving straight field line coordinates.

\subsection{Frieman--Rotenberg formalism}
The formalism introduced by \citet{Frieman_1960} is used here to investigate the stability properties of stationary MHD equilibria. From the full set of linearized MHD equations, 
these authors derived one second-order differential equation for the Lagrangian displacement field vector $\vfg{\xi}$
\begin{equation}
  \vf{G}(\vfg{\xi}) - 2\rho\vf{v}\cdot\grad{\pdb{\vfg{\xi}}{t}} - \rho\frac{\partial^{2}\vfg{\xi}}{\partial t^{2}} = 0,
  \label{eq:friemanrotenberg}
\end{equation}
where the generalized force operator $\vf{G}(\vfg{\xi})$ is
\begin{align}
  \label{eq:forceoperator}
  \vf{G}(\vfg{\xi})                   & = 
    \vf{F}(\vfg{\xi}) + \grad{\Phi}\grad{}\cdot(\rho\vfg{\xi})+ 
    \grad{}\cdot\left[ \rho\vfg{\xi}\vf{v}\cdot\grad{\vf{v}} - \rho\vf{v}\vf{v}\cdot\grad{\vfg{\xi}} \right]. \\
\intertext{Here,}
  \label{eq:forceoperatorstatic}
  \vf{F}(\vfg{\xi}) & = -\grad{\Pi} + \vf{B}\cdot\grad{\vf{Q}} + \vf{Q}\cdot\grad{\vf{B}}
\end{align}
is the force operator for static equilibria without gravity, which was derived by \cite{Bernstein_1958}.
These expressions contain the Eulerian perturbation of the total pressure
\begin{align}
  \label{eq:Etotalpressure}
  \Pi    & = -\gamma p\grad{}\cdot\vfg{\xi} - \vfg{\xi}\cdot\grad{p} + \vf{B}\cdot\vf{Q}, \\
\intertext{where $\gamma$ is the ratio of specific heats, and the Eulerian perturbation of the magnetic field}
  \label{eq:Emagneticfield}
  \vf{Q} & = \grad{} \times \left( \vfg{\xi} \times \vf{B} \right).
\end{align}
The time-dependence of the displacement field is assumed to be exponential with normal mode frequencies~$\omega$, $\vfg{\xi} = \hat{\vfg{\xi}} \exp ( -\im \omega t )$. 
Using this assumption, the Frieman-Rotenberg equation can be written as
\begin{equation}
  \vf{G}(\hat{\vfg{\xi}}) + 2\im \rho\omega\vf{v}\cdot\grad{\hat{\vfg{\xi}}} + \rho\omega^{2}\hat{\vfg{\xi}} = 0.
  \label{eq:FReqn}
\end{equation}
We base our derivations below on the axial symmetry of the solar prominence equilibrium by writing the axial dependence of the displacement field 
as~$\hat{\vfg{\xi}} \sim \exp (\im kz)$, where~$k$ is the axial mode number. We then neglect the hat on~$\hat{\vfg{\xi}}$ and suppress the implied 
$\exp ( -\im \omega t + \im kz)$ dependence for all linear quantities.

\subsection{Field-line projection and representation}
As in all stability studies in fusion research, we need to exploit a projection based on the nested topology of the magnetic surfaces, on which the helical field lines reside. 
This projection ensures that on each flux surface, one can make a distinction between two wave directions: one parallel and the other perpendicular to the magnetic field. 
In the short wavelength limit, these directions correspond to the polarizations of the slow and Alfv\'en waves, respectively. To perform this projection, we adopt straight field 
line, non-orthogonal coordinates $(\psi, \vartheta, z)$ that are introduced in our accompanying paper~\citep{Blokland_2011A}, which are known once the equilibrium is fully 
determined. The triad of unit vectors based on the projection is
\begin{equation}
  \begin{aligned}
    \vf{n}    & \equiv \frac{\grad{\psi}}{|\grad{\psi}|},  & \quad
    \vfg{\pi} & \equiv \vf{b} \times \vf{n},               & \quad
    \vf{b}    & \equiv \frac{\vf{B}}{B},
  \end{aligned}
  \label{eq:basevectors}
\end{equation}
where the gradient of the magnetic flux $\psi$ is perpendicular to the flux surfaces. Using these unit vectors, the components of the displacement 
field~$\vfg{\xi}$ can be replaced by the three related variables
\begin{equation}
  \begin{aligned}
    X & \equiv     \Btheta           \vfg{\xi}\cdot\vf{n},     & \quad
    Y & \equiv \im \frac{B}{\Btheta} \vfg{\xi}\cdot\vfg{\pi},  & \quad
    Z & \equiv \im \frac{1}{B}       \vfg{\xi}\cdot\vf{b},
  \end{aligned}
  \label{eq:xicomponents}
\end{equation}
and the projected gradient operators become
\begin{alignat}{2}
  \label{eq:operatorD}
  \D & \equiv      \frac{1}{\Btheta}\vf{n}   \cdot\grad{}  &  & = \ppsi - \frac{g_{12}}{g_{22}}\ptheta,  \\
  \label{eq:operatorG}
  \G & \equiv -\im \Btheta B        \vfg{\pi}\cdot\grad{}  &  & = \frac{-\im I}{J}\ptheta - k\Btheta^{2},  \\
  \label{eq:operatorF}
  \F & \equiv -\im                  \vf{B}   \cdot\grad{}  &  & = \frac{-\im}{J}\ptheta + \frac{kq}{J}.
\end{alignat}
The second equality in the expressions for $\F$ and $\G$ only holds when these operators act on a
component of the displacement field~$\vfg{\xi}$. These expressions contain several quantities already introduced in paper I, such as the metric elements $g_{ij}$ and the Jacobian $J$ 
for the non-orthogonal straight field line coordinates, and quantities shown to be flux functions to realize equilibrium, such as the poloidal stream function $I=B_z$ and the safety 
factor $q(\psi)=J I$.

The Frieman-Rotenberg equation in Eq.~\eqref{eq:friemanrotenberg} can be written as
\begin{equation}
  \begin{aligned}
    (\ten{A} + \ten{G})\vf{x} - \rho\omt^{2}\ten{B}\vf{x} & = 0,  & 
    \quad \mathrm{with} \quad
    \vf{x} & \equiv \begin{pmatrix} X \\ Y \\ Z \end{pmatrix},
  \end{aligned}
  \label{eq:spectraleqn}
\end{equation}
by exploiting the straight field line coordinates and applying the projection defined by Eq.~\eqref{eq:xicomponents}, where
$\omt(\psi) \equiv \omega - k \vz(\psi)$ is the flux-surface localized Doppler-shifted eigenfrequency. Furthermore, the matrices $\ten{A}$ and~$\ten{B}$ 
are~$3 \times 3$ matrix operators that are also present in the case of a static equilibrium without gravity. The matrix 
operator~$\ten{G}$ contains the elements associated with the external gravitational potential. The non-vanishing matrix elements of~$\ten{A}$ and~$\ten{B}$ are
\begin{equation}
  \begin{aligned}
    \matel{A}_{11} & \equiv -\D\frac{\gamma p + B^{2}}{J}\Dd J + \F\frac{1}{B_{\vartheta}^{2}}\F
                            + 2\left[ \D\left(\Btheta\kp\right) \right],                                            \\
    \matel{A}_{12} & \equiv -\D\gamma p\G \frac{1}{B^{2}} - \D\G 
                            + 2k\Btheta\kp,                                                                          \\
    \matel{A}_{13} & \equiv -\D\gamma p\F,                                                                           \\
    \matel{A}_{21} & \equiv \frac{1}{B^{2}}\G\frac{\gamma p}{J}\Dd J + \G\frac{1}{J}\Dd J 
                            + 2k\Btheta\kp,                                                                          \\
    \matel{A}_{22} & \equiv \frac{1}{B^{2}}\G\gamma p\G\frac{1}{B^{2}} + \G\frac{1}{B^{2}}\G 
                            + \F\frac{\Btheta^{2}}{B^{2}}\F,                                                          \\
    \matel{A}_{23} & \equiv \frac{1}{B^{2}}\G\gamma p\F,                                                             \\
    \matel{A}_{31} & \equiv \F\frac{\gamma p}{J}\Dd J,                                                               \\
    \matel{A}_{32} & \equiv \F\gamma p\G\frac{1}{B^{2}},                                                             \\
    \matel{A}_{33} & \equiv \F\gamma p\F,                                                                            \\
  \end{aligned}
  \label{eq:matrixA}
\end{equation}
and
\begin{equation}
  \begin{aligned}
    \matel{B}_{11} & \equiv \frac{1}{\Btheta^{2}},      &
    \matel{B}_{22} & \equiv \frac{\Btheta^{2}}{B^{2}},  &
    \matel{B}_{33} & \equiv B^{2}.
  \end{aligned}
  \label{eq:matrixB}
\end{equation}
We used the poloidal curvature of the magnetic surfaces $\kp$ from paper I, along with the operators
\begin{equation}
  \begin{aligned}
    \Dd & \equiv \D - \left[ \ptheta\left(\frac{g_{12}}{g_{22}}\right) \right],  &
    \dD & \equiv \D + \left[ \ptheta\left(\frac{g_{12}}{g_{22}}\right) \right].
  \end{aligned}
  \label{eq:operatorsD}
\end{equation}
These are related to the normal gradient operator~$\D$, originally introduced by \citet{Goedbloed_1997} for the spectral analysis of 
static tokamak equilibria. The square brackets in the expressions in Eq.~\eqref{eq:matrixA} for the matrix elements of~$\ten{A}$ and the gradient operators 
defined by \eqref{eq:operatorsD} indicate that the differential operator only acts on the term inside the bracket. This notation is also used in the 
expressions below. As expected, the matrices~$\ten{A}$ and~$\ten{B}$ can also be derived from the expressions for wave modes about static axisymmetric MHD equilibria, found in 
the papers by \citet{Goedbloed_1975,Goedbloed_1997}, by formally setting the toroidal distance $R$ to unity. The matrix~$\ten{G}$ enters if there is an external gravitational 
potential present. The expression for this matrix is 
{\renewcommand{\arraystretch}{2}
\begin{equation}
  \ten{G} =
  \begin{pmatrix}
    -\rho\left\{ \left[ J\dD\left(\dfrac{1}{J}\lambda\right)\right] \right\}                                                   &
      \left( \im\D I\mu + \lambda\G \right) \dfrac{\rho}{B^{2}}                                                                &
      \left( \im\D  \mu + \lambda\F \right) \rho                                                                               \\
    \dfrac{\rho}{B^{2}}\left( \im I\mu\dfrac{1}{J}\Dd J + \G\lambda \right)                                                    &
      -\dfrac{\rho I^{2}}{B^{4}}\left[ \dfrac{1}{J}\ptheta\mu \right]                                                          &
      -\rho\left( \dfrac{I}{B^{2}}\left[ \dfrac{1}{J}\ptheta\mu \right] - \im k \mu \right)                                    \\
    \rho\left( \im\mu\dfrac{1}{J}\Dd J + \F\lambda \right)                                                                     &
      -\rho\left( \dfrac{I}{B^{2}}\left[ \dfrac{1}{J}\ptheta\mu \right] + \im k \mu \right)                                    &
      -\rho\left[ \dfrac{1}{J}\ptheta\mu \right]
  \end{pmatrix},
  \label{eq:matrixR}
\end{equation}
}
where
\begin{align}
  \label{eq:pressuretheta}
  \mu     & \equiv -\frac{1}{J}\ptheta \Phi = \frac{1}{\rho J}\ptheta p  = \frac{1}{\rho}\vf{\Btheta}\cdot\grad{p}, \\
  \lambda & \equiv -\D\Phi.
\end{align}
We note that the term~$\mu$ represents the pressure variation on a flux surface. This matrix $\ten{G}$ can also be derived from the expression for the matrix operator $\ten{R}$ 
found in the paper by \citet{Blokland_2007B}, where axisymmetric, gravitating, magnetized accretion tori with purely toroidal flow where spectroscopically analyzed. The 
equivalence involves (1) setting their radial distance $R=1$, (2) producing the toroidal curvature $\kt=0$ for a translationally symmetric solar prominence, and (3) replacing
the toroidal mode number $n$ by the axial wave number $k$. In that analysis, as well as in the related one for toroidally rotating tokamak plasmas by \citet{vanderHolst_2000B}, 
another matrix $\ten{C}$ related to the Coriolis effect appeared, which naturally vanishes for our translationally invariant configurations.

Another interesting, related limit is the cylindrical limit known from standard textbook treatments~\citep{Book_Goedbloed_2004}. If we ignore gravity altogether, then 
matrix $\ten{G}$ will be zero and the operators $\F$ and $\G$ will just be algebraic expressions. In this case, the spectral equation given in Eq.~\eqref{eq:spectraleqn} 
reduces exactly to the well-known equation presented by \cite{Hain_1958}. They rewrote this matrix form to one second-order differential equation for the radial displacement 
field, assuming a ratio of specific heats $\gamma=1$. This was generalized by \cite{Goedbloed_1971B} for arbitrary $\gamma$, and rewritten, for cases involving 
equilibrium flow in~\cite{Hameiri_1981} and \cite{Bonderson_1987}.  A cylindrical limit involving flow and external gravity, suitable for analyzing accretion disk 
stability, was presented by \cite{Keppens_2002A}, and can also be easily shown to be directly related to the above expressions.

\section{Continuous MHD spectrum \label{sec:continuous}}
In the previous section, we derived the spectral equation in Eq.~\eqref{eq:spectraleqn} governing all MHD waves and instabilities in translational symmetric solar prominences. 
This equation is the starting point to all following MHD spectral computations. In particular, we can derive the equations for the continuous MHD spectrum 
by first considering modes localized on a particular flux surface~$\psi = \psi_{0}$. We can then obtain a non-singular eigenvalue problem in the 
local Alfv\'en and slow-like eigenfunctions $\eta(\vartheta), \zeta(\vartheta)$. This form allows us to derive a stability criterion for continuum modes, where it is 
found that for an axially symmetric prominence, gravity can cause the continuous spectrum to become unstable. By expanding the periodic $\vartheta$-dependence of the 
eigenfunctions, we get an equivalent algebraic set of coupled equations.

\subsection{General formalism}
To derive the equation for the continuous spectrum, we assume that the normal derivative~$(\partial / \partial \psi)$ of the 
eigenfunctions is large compared to the eigenfunctions themselves and the local Doppler-shifted eigenfrequency~$\omt (\psi)$ is finite. 
Using these assumptions, the first row of the spectral equation in Eq.~\eqref{eq:spectraleqn} can be solved approximately
\begin{equation}
  \frac{1}{J}\Dd JX \approx \frac{-\gamma p}{\gamma p + B^{2}} \left( \G \frac{1}{B^{2}} Y + \F Z \right)
                            - \frac{1}{\gamma p + B^{2}}\G Y
			    + \frac{\im \rho\mu}{\gamma p + B^{2}} \left( \frac{I}{B^{2}}Y + Z \right).
  \label{eq:approxX}
\end{equation}
We note that $X$ is small compared to either $Y$ or $Z$. This implies that the continuum modes are mainly tangential to
a particular flux surface. Furthermore, we note that $\partial X / \partial \psi$, $Y$, and $Z$ are of the same order.

Inserting the approximated solution given by Eq.~\eqref{eq:approxX} into the second and third row of the spectral equation in Eq.~\eqref{eq:spectraleqn}
results in an eigenvalue problem that is independent of the normal derivative. Hence, the reduced problem becomes non-singular. Exploiting this property, 
we can write the projected displacement field components~$Y$ and $Z$ as
\begin{equation}
  \begin{aligned}
    Y & \approx \delta (\psi - \psi_{0}) \eta (\vartheta), \\
    Z & \approx \delta (\psi - \psi_{0}) \zeta(\vartheta),
  \end{aligned}
\end{equation}
where $\delta (\psi - \psi_{0})$ is the Dirac delta function. Here, $Y$ and $Z$ have been divided into an improper ($\psi$--dependence) and 
proper part ($\vartheta$--dependence). This kind of splitting was introduced by \citet{Goedbloed_1975} for static axisymmetric 
equilibria without gravity. The reduced, non-singular eigenvalue problem becomes
\begin{equation}
  \left( \ten{a} + \ten{g} - \rho\omt^{2}\ten{b} \right)
  \begin{pmatrix} \eta \\ \zeta \end{pmatrix} = 0,
  \label{eq:continuousspectrum}
\end{equation}
where
{\renewcommand{\arraystretch}{2}
\begin{equation}
  \ten{a} = 
  \begin{pmatrix}
    \F\dfrac{\Btheta^{2}}{B^{2}}\F + 4\dfrac{\gamma p}{\gamma p + B^{2}} \left( \Btheta\kg \right)^{2}  &
      -2\im\dfrac{\gamma pB}{\gamma p + B^{2}} \left( \Btheta\kg \right)\F                              \\
    2\im\F\dfrac{\gamma pB}{\gamma p + B^{2}}\left( \Btheta\kg \right)                                  &
      \F\dfrac{\gamma pB^{2}}{\gamma p + B^{2}}\F
  \end{pmatrix},
  \label{eq:matrixa}
\end{equation}
}
{\renewcommand{\arraystretch}{2}
\begin{equation}
  \ten{g} =
  \begin{pmatrix}
    -4\dfrac{I}{B^{2}}\dfrac{\rho\mu B}{\gamma p + B^{2}} \left( \Btheta\kg \right) + \dfrac{\rho I^{2}}{B^{2}}\Nmpol                   &
      \im I\dfrac{\rho\mu B}{\gamma p + B^{2}} B^{2}\F\dfrac{1}{B^{2}} + \rho I \Nmpol                                                  \\
    -\im\dfrac{I}{B^{2}}\F\dfrac{\rho\mu}{\gamma p + B^{2}} B^{2} + \rho I \Nmpol                                                       &
      -B^{2}\left[ \dfrac{1}{JB^{2}}\ptheta\dfrac{\rho\mu}{\gamma p + B^{2}} B^{2} \right] + \rho B^{2} \Nmpol
  \end{pmatrix},
  \label{eq:matrixr}
\end{equation}
}
{\renewcommand{\arraystretch}{2}
\begin{equation}
  \ten{b} =
  \begin{pmatrix}
    \dfrac{\Btheta^{2}}{B^{2}}  &
      0                         \\
    0                           &
      B^{2}
  \end{pmatrix}.
  \label{eq:matrixb}
\end{equation}
}
In the expressions above, $\kg$ is the geodesic curvature,
\begin{equation}
    \kg = \vfg{\kappa}\cdot\vfg{\pi}          
        = \frac{I}{J\Btheta B^{2}}\ptheta B,
\end{equation}
where $\vfg{\kappa} = \vf{b}\cdot\grad{\vf{b}}$ is the field line curvature. Furthermore, $N_{\mathrm{m,pol}}^{2}$ is the magnetically modified 
Brunt-V\"ais\"al\"a frequency projected onto a flux surface,
\begin{equation}
  \begin{aligned}
    N_{\mathrm{m,pol}}^{2} & = \frac{\mu}{B^{2}}\left[ \frac{1}{J\rho}\ptheta\rho - \frac{\rho\mu}{\gamma p + B^{2}} \right] \\
                           & = \left[ \frac{\vf{\Btheta}\cdot\grad{p}}{\rho B} \right] 
                               \left[ \frac{\vf{\Btheta}}{\rho B}\cdot\left( \grad{\rho}  - \frac{\rho}{\gamma p + B^{2}} \grad{p} \right) \right].
  \end{aligned}
\end{equation}
This magnetically modified Brunt-V\"ais\"al\"a frequency has also been found in various related cases. For example, \citet{Poedts_1985} and \citet{Belien_1997}
found this frequency for static, axisymmetric gravitationally stratified plasmas. \citet{vanderHolst_2000A,vanderHolst_2000B} discussed its role for tokamak plasmas with purely 
toroidal flow. \citet{Blokland_2007B} generalized the expression by including an arbitrary external gravitational potential in its application to rotating, magnetized accretion 
disks. When we examine the matrix $\ten{g}$ closely, we see that the poloidal variation in the pressure is present in all its matrix elements. This has the consequence that the 
diagonal dominance of matrix $\ten{a}$ will be destroyed, which causes possible mode coupling between Alfv\'en and slow modes.

The non-singular eigenvalue problem presented in Eq.~\eqref{eq:continuousspectrum} is solved for each flux surface, using poloidally periodic boundary conditions for the 
eigenfunctions~$\eta$ and $\zeta$. Typically, one solves the problem for a given axial mode number $k$, and we then get a set of discrete eigenvalues for each flux surface. 
When we quantify these eigenfrequencies for all flux surfaces, they together map out the continuous spectra. For completeness, we point out that if one uses a formulation based 
on the primitive variables $\rho$, $\vf{v}$, $p$, and $\vf{B}$, an additional entropy continuum $kv_{z}$ is found. \citet{Goedbloed_2004B} showed that this Eulerian entropy 
continuum, in ideal MHD, does not couple to any of the other continua, nor to the discrete stable and/or unstable modes.

\subsection{Spectral properties and stability criterion}
We now derive a general stability criterion for the continua. For this derivation, a Hilbert space with appropriate inner product
needs to be constructed. The parallel gradient operator~$\F$ is a Hermitian operator under the inner product
\begin{equation}
  \oint J v^{*} \left( \F w \right) \dd{}\vartheta = \oint J \left( \F v \right)^{*} w \dd{}\vartheta
  \label{eq:HermitianF}
\end{equation}
for poloidally periodic functions $v$ and $w$. Inspired by this property, a Hilbert space can be defined with the inner product
\begin{equation}
  \langle \vf{v},\vf{w} \rangle \equiv \oint \vf{v}^{*}\cdot\vf{w} J \dd{}\vartheta .
  \label{eq:innerproduct}
\end{equation}
In this Hilbert space, the matrix operators~$\ten{a}$, $\ten{b}$, and $\ten{g}$ are Hermitian operators. Setting~$\vf{v}$ and~$\vf{w}$ equal 
to the eigenfunctions~$(\eta,\zeta)^{\mathrm{T}}$, one can derive a simple quadratic polynomial for the eigenfrequency~$\omt$ from the spectral 
equation in Eq.~\eqref{eq:continuousspectrum} for the continua, namely
\begin{equation}
  a\omt^{2} - c = 0,
  \label{eq:polynomial}
\end{equation}
where the coefficients are
\begin{align}
  \label{eq:ointa}
  a  & \equiv \oint \rho\left( \frac{\Btheta^{2}}{B^{2}}|\eta|^{2} + B^{2}|\zeta|^{2} \right) J\dd{}\vartheta,  \\
  \label{eq:ointc}
  c  & \equiv \oint \left\{   \frac{\Btheta^{2}}{B^{2}} \left| \F\eta \right|^{2}
                            + \frac{\gamma pB^{2}}{\gamma p + B^{2}} \left| \F\zeta + 2\im\frac{\Btheta\kg}{B}\eta - \im\frac{\rho\mu}{\gamma p}\left(\frac{I}{B^{2}}\eta + \zeta \right) \right|^{2} \right. \\
     & \phantom{\equiv \oint \left\{\right.} \left.
	   		    + \rho B^{2}N_{\mathrm{BV,pol}}^{2} \left| \frac{I}{B^{2}}\eta + \zeta \right|^{2} \right\} J\dd{}\vartheta \nonumber.
\end{align}
Here, $N_{\mathrm{BV,pol}}^{2}$ is the Brunt-V\"ais\"al\"a frequency projected onto a flux surface
\begin{equation}
  \begin{aligned}
    N_{\mathrm{BV,pol}}^{2} & = \frac{\mu}{B^{2}}\left[ \frac{1}{J\rho}\ptheta\rho - \frac{\rho\mu}{\gamma p} \right]  \\
                         & =  \left[ \frac{\vf{\Btheta}\cdot\grad{p}}{\rho B} \right] 
                              \left[ \frac{\vf{\Btheta}}{\rho B}\cdot\left( \grad{\rho}  - \frac{\rho}{\gamma p} \grad{p} \right) \right] \\
			 & = -\left[ \frac{\vf{\Btheta}\cdot\grad{p}}{\rho B}    \right]
                              \left[ \frac{\vf{\Btheta}\cdot\grad{S}}{\gamma BS} \right].
  \end{aligned}
\end{equation}
The coefficient $a$ is real and always greater than zero. As for $a$, the coefficient $c$ is real but can be zero or even negative. The solutions of this second order 
polynomial in Eq.~\eqref{eq:polynomial} are simply $\omt = \pm \sqrt{c/a}$. If $c$ is negative, a damped stable wave and an overstable mode are obtained. In this case, 
its absolute value is found from $|\omt|=\sqrt{-c / a}$, while $\mathrm{Re}(\omega) = kv_{z}$, so the oscillation frequency is set by the local Doppler shift.
By looking at the expression for $c$, we realize that this coefficient is equal to or greater than zero if
\begin{equation}
  N_{\mathrm{BV,pol}}^{2} \ge 0
  \label{eq:stabilitycriterion}
\end{equation}
is satisfied everywhere in the plasma. Therefore, equilibria for which the entropy or the temperature (assuming $\gamma > 1$) is a flux function, will always be stable according to 
this criterion. However, when density is a flux function, the equilibrium variation may violate this criterion. This implies that such equilibria have unstable MHD continua, 
a situation identified in earlier, related work, as being caused by the convective continuum instability (CCI) \citep{vanderHolst_2000A,vanderHolst_2000B,Blokland_2007B}. 
This CCI instability criterion in Eq.~\eqref{eq:stabilitycriterion} is analogous to the Schwarzschild criterion for convective instability, with one important difference. 
The Schwarzschild criterion deals with normal derivatives, while this criterion deals with tangential ones. This difference has also been found by various 
authors (\cite{Hellsten_1979,Hameiri_1983,Poedts_1985,vanderHolst_2000B,Blokland_2007B}).

The relevance of this CCI instability to solar prominences may well be significant. We note that it is common practice to study the (slow) evolution of prominences carrying 
flux ropes by taking an equilibrium configuration, and subjecting it to e.g. photospheric-boundary-driven changes, and relax it thereby repeatedly to successive equilibrium 
states~\citep[see e.g.][]{Amari99}. As a result, the internal flux rope variation proceeds through a series of equilibria. For short-lived prominences, equilibria with 
density as a flux function may occur, and in this evolutionary process, the equilibrium can transit from the stable to the unstable regime as governed by the CCI criterion. As 
we later demonstrate, the CCI growth rate also varies with increasing gravity parameter $g$. As explained in Paper I, this $g$-increase could relate to an increase in size, 
internal density, or the weaker overall magnetic field strength of the prominence carrying flux rope.

\subsection{Expansion in poloidal harmonics}
We now continue to transform the two coupled differential equations for the MHD continua in Eq.~\eqref{eq:continuousspectrum} to an algebraic set,
by expanding the eigenfunctions $\eta$ and $\zeta$ in poloidal Fourier harmonics. This expansion has the advantage that we can perform
a detailed investigation of how different poloidal harmonics may couple to each other. This coupling could either be between modes of the same kind, 
for example Alfv\'en-Alfv\'en coupling, or between modes of different kind such as Alfv\'en-slow coupling. In the next section, we show that in an actual solar 
prominence, more complex mode coupling is present.

Expanding the eigenfunctions $\eta$ and $\zeta$ for the MHD continua in a finite Fourier series,
\begin{equation}
  \begin{pmatrix}
    \eta  \\
    \zeta
  \end{pmatrix}
  =
  \sum_{m'}
  \begin{pmatrix}
    \eta_{m'}  \\
    \zeta_{m'}
  \end{pmatrix}
  \e^{\im m'\vartheta}
  \,,
  \label{eq:eta_zeta_harmonics} 
\end{equation}
and applying the operator
\begin{equation}
  \oint J\e^{-\im m\vartheta}\dd{}\vartheta
  \label{eq:operator_m}
\end{equation}
to the spectral equation in Eq.~\eqref{eq:continuousspectrum}, one obtains the algebraic coupled equations
\begin{equation}
  \begin{pmatrix}
    K_{mm'} & M_{mm'}  \\
    N_{mm'} & L_{mm'}  
  \end{pmatrix}
  \begin{pmatrix}
    \eta_{m'}  \\
    \zeta_{m'} 
  \end{pmatrix}
  = 0.
  \label{eq:continuousspectrum_m}
\end{equation}
The operator in Eq.~\eqref{eq:operator_m} is the result of exploiting the inner product in Eq.~\eqref{eq:innerproduct} of the previously defined Hilbert space.
The poloidal mode numbers $m$ and $m'$ have the same range and the $2 \times 2$ submatrices are given by
\begin{equation}
  \sum_{m'}
  \begin{pmatrix}
    K_{mm'} & M_{mm'}  \\
    N_{mm'} & L_{mm'}
  \end{pmatrix}
  = \frac{1}{2\pi} \oint
  \begin{pmatrix}
    K(\vartheta) & M(\vartheta)  \\
    N(\vartheta) & L(\vartheta)
  \end{pmatrix}
  \e^{-\im (m'-m)\vartheta} \ \dd{}\vartheta.
  \label{eq:submatrices}
\end{equation}
The explicit expressions for the integrand functions $K(\vartheta)$, $L(\vartheta)$, $M(\vartheta)$, and $N(\vartheta)$ are
\begin{align}
  \label{eq:elementK}
  K(\vartheta) & = J\frac{\Btheta^{2}}{B^{2}}
                   \left[    \frac{(m+kq)(m'+kq)}{J^{2}} + 4\frac{\gp B^{2}}{\gpB}\kg^{2} - \rho\omt^{2}
                          -4 \frac{I}{\Btheta}\frac{\rho\mu B}{\gpB}\kg + \frac{I^{2}}{\Btheta^{2}}\rho\Nmpol \right],                               \\
  \label{eq:elementM}
  M(\vartheta) & = J\Btheta
                   \left[   -2\im\frac{\gp B}{\gpB}\kg\frac{m'+kq}{J}
                          + \frac{\rho\mu}{\gpB} \left(  \im\frac{I}{\Btheta}\frac{m'+kq}{J} - 2B\kg \right) + \frac{I}{\Btheta}\rho\Nmpol \right],  \\
  \label{eq:elementN}
  N(\vartheta) & = J\Btheta
                   \left[   2\im\frac{\gp B}{\gpB}\kg\frac{m'+kq}{J}
                          + \frac{\rho\mu}{\gpB} \left( -\im\frac{I}{\Btheta}\frac{m'+kq}{J} - 2B\kg \right) + \frac{I}{\Btheta}\rho\Nmpol \right],  \\
  \label{eq:elementL}
  L(\vartheta) & = JB^{2}
                   \left[  \frac{\gp}{\gpB}\frac{(m+kq)(m'+kq)}{J^{2}} - \rho\omt^{2}
                          -\im\frac{\rho\mu}{\gpB}\frac{m-m'}{J} + \rho\Nmpol           \right].
\end{align}
These expressions are derived by exploiting the Hermitian property in Eq.~\eqref{eq:HermitianF} of the parallel gradient operator $\F$.

Once again, we consider the cylindrical limit in combination with zero gravity, which is appropriate for many earlier seismological studies of prominences. In this limit, 
all equilibrium quantities are independent of the poloidal angle $\vartheta$, therefore the geodesic curvature will be zero. This means that the Alfv\'en and slow continuum 
decouple fully from each other. Furthermore, only one single harmonic is required for the expansion of the Fourier series of the eigenfunctions $\eta$ and $\zeta$. Solving the 
spectral equation for this case, one finds the well-known local Doppler-shifted Alfv\'en continua
\begin{align}
  \label{eq:w_alfven}
  \Omega_{\mathrm{A},m}^{\pm} & = kv_{z} \pm \omega_{\mathrm{A},m},  &  \omega_{\mathrm{A},m} & = \left(\frac{m}{r}B_{\vartheta} + kB_{z}\right)/\sqrt{\rho}, \\
\intertext{and the local Doppler-shifted slow continua}
  \label{eq:w_slow}
  \Omega_{\mathrm{s},m}^{\pm} & = kv_{z} \pm \omega_{\mathrm{s},m},  &  \omega_{\mathrm{s},m} & = \sqrt{\frac{\gp}{\gpB}}\ \omega_{\mathrm{A},m}.
\end{align}
Looking at the expressions in Eqs.~\eqref{eq:elementK}-\eqref{eq:elementL} for the submatrix elements, we can make some general statements about the 
MHD continua of solar prominences. In the absence of gravity ($\mu = \Nmpol = 0$), the Alfv\'en and slow continuum can be coupled by the geodesic 
curvature present in the off-diagonal terms $M(\vartheta)$ and $N(\vartheta)$. With gravity, these off-diagonal terms can become more pronounced. 
The second statement is that an Alfv\'en continuum at a resonant surface, associated with the element $K(\vartheta)$, may shift away from the local 
Doppler-shifted frequency $kv_{z}$ because of the (1) geodesic curvature, involving its square $\kg^{2}$, (2) compressibility and finite pressure ($\gamma p$), 
and (3) the pressure variation on a flux surface. All these effects suggest that an Alfv\'en gap can appear around the local Doppler-shifted frequency 
and that Alfv\'en-slow coupling should be taken into account for these gaps. This kind of coupling is investigated in the next section.

\section{Small gravity expansion}\label{sec:expand}
In previous sections, we have derived equations for the continuous MHD spectrum of a translationally symmetric equilibrium. These derivations were completely general, 
and allowed us to identify an instability criterion on the one hand, as well as to identify the terms responsible for Alfv\'en-slow type coupling. In this section, we 
quantify this coupling effect, using the small gravity parameter expansion for the equilibrium, similar to those discussed in paper I. In the remainder of this paper, we assume 
a gravitational potential in which the prominence is embedded given by
\begin{equation}
  \Phi(x,y) = (x-x_{0})g,
  \label{eq:gravity}
\end{equation}
where $x_{0}$ is the location of the center of the last closed flux surface of the prominence and the role of gravity is represented by the constant $g$.
This gravity parameter $g$ denotes a dimensionless quantification of the relative importance of gravity on the equilibrium structure, and we explained in paper I that this parameter 
can meaningfully vary by orders of magnitude for solar prominence configurations. For small $g$ and assuming a circular cross-section of the prominence plasma, we quantified in 
paper I the change induced in the equilibrium (the downward shift of the flux surfaces), and the corresponding straight field line coordinates. This kind of expansion is similar to the 
small inverse aspect ratio expansion for tokamak plasmas. Using this analytical result for the equilibrium, we can now demonstrate that gaps (or even instabilities) may appear in the 
continuous MHD spectrum because of mode coupling, as a direct consequence of the force of gravity. 

\subsection{Continuous MHD spectrum for equilibria weakly affected by gravity}
We apply the small gravity expansion to the expressions found in Eqs.~\eqref{eq:elementK}, \eqref{eq:elementM}, \eqref{eq:elementN}, and \eqref{eq:elementL} for the matrix 
elements $K(\vartheta)$, $M(\vartheta)$, $N(\vartheta)$, and $L(\vartheta)$, respectively. For this expansion, we rely on the metric elements that are valid for straight 
field line coordinates exploiting the Shafranov shift $\Delta(r)$, given by Eq.~(26) in our accompanying paper \citep{Blokland_2011A}. In addition, one can derive the 
quantities
\begin{alignat}{2}
  \label{eq:smallg_rho}
  \rho                 & \approx \rho_{0} \left[ 1 - t_{e} \frac{r\rho_{0}g}{p_{0}}\cos(\vartheta)                  \right]   & & = \mcO(1), \\
  \label{eq:smallg_p}
  p                    & \approx p_{0}    \left[ 1 - \frac{r\rho_{0}g}{p_{0}}\cos(\vartheta)                        \right]   & & = \mcO(1), \\
  \label{eq:smallg_Bpol}
  \Btheta              & \approx \psi'    \left[ 1 + \Delta'\cos (\vartheta)                                        \right]   & & = \mcO(1), \\
  \label{eq:smallg_kg}
  \kg                  & \approx -t_{g}\frac{{\psi'}^{2}}{B_{0}^{2}}\frac{I}{B_{0}}\frac{\Delta'}{r}\sin(\vartheta)           & & = \mcO(g), \\
  \label{eq:smallg_mu}
  \frac{\rho\mu}{\gpB} & \approx \frac{\rho_{0}g}{\gpBz} \psi'\sin(\vartheta)                                                 & & = \mcO(g), \\
  \label{eq:smallg_Nmpol}
  \rho\Nmpol           & \approx \frac{{\psi'}^{2}}{B_{0}^{2}} \left( \rho_{0} g \right)^{2} 
                                 \left( \frac{t_{e}}{p_{0}} - \frac{1}{\gamma p_{0} + B_{0}^{2}} \right) \sin^{2}(\vartheta)  & & = \mcO(g^{2}),
\end{alignat}
where the total magnetic field $B_{0}^{2} = {\psi'}^{2} + I^{2}$. The parameter $t_{g}=1$ has been introduced to track the linear and quadratic dependence on the geodesic curvature. 
Furthermore, the parameter $t_{e}$ allows us to distinguish the three options for the flux function dependence in the equilibrium, namely
\begin{equation}
  t_{e} =
  \begin{cases}
    1         & \text{for } T(\psi),    \\
    0         & \text{for } \rho(\psi), \\
    1/\gamma  & \text{for } S(\psi).
  \end{cases}
\end{equation}
These flux functions lead to different variants of the extended Grad-Shafranov equation in Eq.~(10) for the equilibrium, as presented in our accompanying paper \citep{Blokland_2011A}. 
We again note that the magnetically modified Brunt-V\"ais\"al\"a frequency $\Nmpol$ from Eq.~\eqref{eq:smallg_Nmpol} can only be negative if the density is a flux function. This once 
more confirms that for this particular choice of the flux function, the MHD continua may become unstable as already found to be true in general above.

Applying the small gravity expansion, the eigenvalue equation in Eq.~\eqref{eq:continuousspectrum_m} can be written as
\begin{equation}
  \sum_{m'}
  \begin{pmatrix}
    \overline{K}_{mm'} & \overline{M}_{mm'}  \\
    \overline{N}_{mm'} & \overline{L}_{mm'}  
  \end{pmatrix}
  \begin{pmatrix}
    \overline{\eta}_{m'}  \\
    \overline{\zeta}_{m'} 
  \end{pmatrix}
  = 0,
  \label{eq:smallg_continuousspectrum_m}
\end{equation}
where
\begin{align}
  \overline{K}_{mm'}      & \equiv \frac{qB_{0}^{2}}{r^{2}I} K_{mm'},    &
    \overline{M}_{mm'}    & \equiv \frac{1}{r} M_{mm'},                  &
    \overline{\eta}_{m'}  & \equiv \frac{r}{q} \eta_{m'},                \\
  \overline{N}_{mm'}      & \equiv \frac{1}{r} N_{mm'},                  &
    \overline{L}_{mm'}    & \equiv \frac{I}{qB_{0}^{2}} L_{mm'},         &
    \overline{\zeta}_{m'} & \equiv \frac{B_{0}^{2}}{I} \zeta_{m'}.
\end{align}
We note that the first and second rows of the eigenvalue equation in Eq.~\eqref{eq:smallg_continuousspectrum_m} have been multiplied by $B_{0}^{2} / rI$ and $q^{-1}$, 
respectively, to get rid of the dimensional factors in front of the square brackets in the definitions given in Eqs.~\eqref{eq:elementK}-\eqref{eq:elementL}. The approximated 
expressions of the integrands of the matrix elements used in the spectral equation in Eq.~\eqref{eq:smallg_continuousspectrum_m} are now
\begin{align}
  \label{eq:smallg_elementK}
  \overline{K}(\vartheta) & = \frac{I^{2}}{q^{2}}(m+kq)(m'+kq) - \rho_{0}\omt^{2}                                                                                                                                           \\
                          & \quad
                              + 2\frac{I^{2}}{B_{0}^{2}}\left[ \frac{I^{2}}{q^{2}}(m+kq)(m'+kq)\Delta' - \rho_{0}\left( \Delta' - t_{e}\frac{B_{0}^{2}}{I^{2}}\frac{r\rho_{0}g}{p_{0}}\right) \omt^{2} \right] \cos(\vartheta)  \nonumber \\
                          & \quad
                              + 4\frac{I^{2}}{\gpz (\gpBz)} t_{g}\alpha \left( t_{g}\alpha + \rho_{0}g \right) \sin^{2}(\vartheta) + \frac{q^{2}}{r^{2}}\orNmpol\sin^{2}(\vartheta),                                        \nonumber \\
  \label{eq:smallg_elementM}
  \overline{M}(\vartheta) & = \im\frac{I^{2}}{q}\frac{m'+kq}{\gpBz} \left( 2t_{g}\alpha + \rho_{0}g \right) \sin (\vartheta)
                              + 2t_{g}\frac{r^{2}I^{4}}{q^{3}B_{0}^{2}} \frac{\rho_{0}g\Delta'}{\gpBz}\sin^{2}(\vartheta) + \frac{q}{r}\orNmpol\sin^{2}(\vartheta),                                                         \\
  \label{eq:smallg_elementN}
  \overline{N}(\vartheta) & = -\im\frac{I^{2}}{q}\frac{m+kq}{\gpBz} \left( 2t_{g}\alpha + \rho_{0}g \right) \sin (\vartheta)
                              + 2t_{g}\frac{r^{2}I^{4}}{q^{3}B_{0}^{2}} \frac{\rho_{0}g\Delta'}{\gpBz}\sin^{2}(\vartheta) + \frac{q}{r}\orNmpol\sin^{2}(\vartheta),                                                         \\
  \label{eq:smallg_elementL}
  \overline{L}(\vartheta) & = \frac{\gpz}{\gpBz}\frac{I^{2}}{q^{2}}(m+kq)(m'+kq) - \rho_{0}\omt^{2}                                                                                                                         \\
                          & \quad
                              + \left[ \frac{\gpz}{\gpBz}\frac{rI^{2}}{q^{2}(\gpBz)}\left( 2\alpha - \frac{B_{0}^{2}}{p_{0}} \rho_{0}g \right) (m+kq)(m'+kq)
                                        - \frac{r\rho_{0}}{B_{0}^{2}}\left( 2\frac{I^{2}}{q^{2}}r\Delta' - t_{e}\frac{B_{0}^{2}}{p_{0}}\rho_{0}g \right)
                                \right] \cos (\vartheta) \nonumber \\
                          & \quad
                              - \im \frac{r\rho_{0}g}{\gpBz}\frac{I^{2}}{q^{2}}(m-m') \sin(\vartheta) + \orNmpol\sin^{2}(\vartheta),  \nonumber          \\
\intertext{where}
  \label{eq:smallg_alpha}
  \alpha                  & \equiv \frac{\gpz}{B_{0}^{2}}\frac{I^{2}}{q^{2}}r\Delta', \\
  \label{eq:smallg_oNmpol}
  \orNmpol                & \equiv \frac{r^{2}I^{2}}{q^{2}B_{0}^{2}} \left( \rho_{0}g \right)^{2} \left( \frac{t_{e}}{p_{0}} - \frac{1}{\gpBz} \right).
\end{align}
Once more, the magnetically modified Brunt-V\"ais\"al\"a frequency $\orNmpol$ \eqref{eq:smallg_oNmpol} is only negative if the density is a 
flux function ($t_{e}=0$). 

\subsection{Low frequency $\Delta m=0$ gap}
The approximate relations just given can now be analyzed in terms of mode coupling possibilities. We focus on low frequencies, $\omt = \mcO(g)$, and show that a 
$\Delta m = 0$ gap can be created in the neighborhood of a rational surface $m+kq=0$, where $q$ quantifies the flux-surface specific local twist in the field lines on this surface. 
For such a surface, we determine the leading order coupling on an Alfv\'en continuum mode $\overline{\eta}_{m}$ with the same poloidal mode number $m$. Close to the rational surface, 
we can safety neglect the third term of the expression in Eq.~\eqref{eq:smallg_elementK} for $\overline{K}$ and Eq.~\eqref{eq:smallg_elementL} for $\overline{L}$. We are left with 
the eigenvalue equation for the MHD continua in Eq.~\eqref{eq:smallg_continuousspectrum_m}, which reduces in essence to a four-mode coupling between the central Alfv\'en mode, and the 
three neighboring slow modes given by
\begin{equation}
  \begin{pmatrix}
    \rho_{0}\hat{\omega}_{\mathrm{s},m-1}^{2} - \rho_{0}\omt^{2}  
      & b_{m-1} 
      & c_{m-1,m} 
      & d         \\
    b_{m-1}
      & \rho_{0}\hat{\omega}_{\mathrm{A},m}^{2}  - \rho_{0}\omt^{2}
      & e        
      & -b_{m+1}  \\
    c_{m-1,m} 
      & e 
      & \rho_{0}\hat{\omega}_{\mathrm{s},m}^{2} - \rho_{0}\omt^{2}
      & c_{m,m+1}  \\
    d
      & -b_{m+1}
      & c_{m,m+1}
      & \rho_{0}\omega_{\mathrm{s},m+1}^{2} - \rho_{0}\omt^{2}
  \end{pmatrix}
  \begin{pmatrix}
    \overline{\zeta}_{m-1} \\
    \overline{\eta}_{m}    \\
    \overline{\zeta}_{m}   \\
    \overline{\zeta}_{m+1}
  \end{pmatrix}
  = 0,
  \label{eq:smallg_deltam}
\end{equation}
where
\begin{align}
  \rho_{0}\hat{\omega}_{\mathrm{s},m}^{2} & \equiv \rho_{0}\omega_{\mathrm{s},m}^{2} + \frac{1}{2}\orNmpol,  \\
  \rho_{0}\hat{\omega}_{\mathrm{A},m}^{2} & \equiv \rho_{0}\omega_{\mathrm{A},m}^{2} + 2\frac{I}{\gp (\gpB)}t_{g}\alpha\left( t_{g}\alpha + \rho_{0}g \right) + \frac{q^{2}}{2r^{2}}\orNmpol,  \\
  b_{m}                                   & \equiv \frac{I}{2(\gpB)} \left( 2t_{g}\alpha + \rho_{0}g \right) \sqrt{\rho_{0}}\omega_{\mathrm{A},m},  \\
  c_{m,m'}                                & \equiv \frac{r}{2(\gpB)} \left( 2\alpha - \frac{B_{0}^{2}}{p_{0}}\rho_{0}g \right)\rho_{0}\omega_{\mathrm{s},m}\omega_{\mathrm{s},m'} - \frac{r}{2(\gpB)}\frac{I^{2}}{q^{2}}\rho_{0}g, \\
  d                                       & \equiv -\frac{1}{4}\orNmpol, \\
  e                                       & \equiv \frac{r}{q} \left( t_{g}\alpha \frac{I^{2}}{\gp (\gpB)}\rho_{0}g + \frac{1}{2}\frac{q^{2}}{r^{2}}\orNmpol \right).
\end{align}
In this derivation, we made use of the finding that the integral of $\sin(\vartheta)$ and $\sin^{2}(\vartheta)$ over $\vartheta$ gives 
$\tfrac{1}{2}\im(\delta_{m,m'+1} + \delta_{m,m'-1})$ and $-\tfrac{1}{4}(\delta_{m,m'-2} - 2\delta_{m,m'} + \delta_{m,m'+2})$, respectively.
We note that for this four-mode coupling scheme, we neglected any coupling to the $m \pm 1$ Alfv\'en modes, an assumption that is valid as long 
as the associated frequencies are far away in the spectrum. The same argument also holds for the coupling to the $m \pm 2$ Alfv\'en and slow modes. 
By careful examination of this $4\times 4$ matrix, one notices that the coupling between $\overline{\zeta}_{m \pm 1}$ with either $\overline{\eta}_{m}$ 
or $\overline{\zeta}_{m}$ is on the order of $\mcO(g)$, while all the other types of coupling are on the order of $\mcO(g^{2})$.
Because the main modes involved have the same poloidal mode number $m$, this coupling is known as a $\Delta m=0$ gap possibility.

The matrix of the reduced eigenvalue problem presented in Eq.~\eqref{eq:smallg_deltam} is symmetric and therefore its eigenvalues $\omt^{2}$ will be real. This is consistent with 
the properties of the quadratic polynomial  \eqref{eq:polynomial} for the eigenfrequency $\omt$. If gravity drives a certain MHD continuum unstable, it means that the corresponding 
eigenfrequency $\omt^{2}$ will be smaller than zero, resulting in a complex eigenvalue $\omt$. Once the eigenvalue $\omt$ is known, it is straightforward to determine
the dominant character of the MHD mode by determining the dominant component of the eigenvector of the eigenvalue equation in Eq.~\eqref{eq:smallg_deltam}. This analytic result will 
be used in what follows to verify the numerically obtained continua.

\section{The spectral code PHOENIX}\label{sec:phoenix}
While the equilibria presented in paper I were computed with the FINESSE code~\citep{Belien_2002}, the stability analysis of the prominence models is subsequently done using the 
spectral code PHOENIX~\citep{Blokland_2007A}. Although this code was originally developed for analyzing axisymmetric plasma tori such as tokamaks and accretion disks,
the design is flexible enough to analyze axially symmetric, gravitating plasma tubes. This possibility is exploited in the present PHOENIX version 1.2.
For a given two-dimensional equilibrium computed by FINESSE and converted to straight field line coordinates, we then need to select an axial wave number $k$ and a range of 
poloidal wave numbers $m$ for which we determine the complete MHD spectrum numerically. A range of poloidal wave numbers is required to allow for the anticipated mode coupling 
caused by the combinations of non-circular cross-sections of the flux surfaces and/or the presence of a gravitational potential. Instead of solving the spectral equation 
in Eq.~\eqref{eq:spectraleqn}, PHOENIX solves the linearized version of the full set of non-linear MHD equations. This allows us to extend the eigenspectrum computations for 
future incorporation of non-adiabatic effects, while it in addition will contain the trivial solution for the entropy modes known as the Eulerian entropy continuum. The 
discretization of these linearized MHD equations exploits a finite element method in the radial direction, while in the poloidal direction a spectral method is applied. The 
finite elements used are a combination of quadratic and cubic Hermite elements. This combination prevents the creation of spurious eigenvalues~\citep{Ra}. The resulting 
generalized non-Hermitian eigenvalue problem is then solved using the Jacobi-Davidson method~\citep{Sleijpen_1996}. For such an eigenspectrum computation, one needs to supply 
boundary conditions at both the magnetic axis (set by symmetry) and the outer flux surface. There, we assume a perfectly conducting wall for simplicity. Strictly speaking, 
this condition is not applicable to true flux ropes carrying embedded solar prominences. However, since we focus on the continuous spectrum alone, the boundary conditions in 
the radial direction have hardly any influence, since those waves or instabilities are by definition localized on flux surfaces. 

Exploiting this realization, \cite{Poedts_1993} introduced a more efficient method than the Jacobi-Davidson method to compute the continuous MHD spectrum. This method 
replaces the radial dependence of the cubic Hermite and the quadratic elements with $\log(\epsilon)$ and $1/\epsilon$, respectively, on each individual flux surface. In this way, 
the singular behavior of the continuous modes can be approximated. \citet{Goedbloed_1975} and \citet{Pao_1975} describe in detail the reasons for this type of singular behavior 
in the continua for static plasma tori without gravity. \cite{Goedbloed_2004A} pointed out that similarly singular behavior for the MHD continua prevails in plasma tori including 
toroidal plus poloidal flows and gravity. The same behavior is thus also expected for our translationally symmetric case. After replacing the elements, the eigenvalue problem 
becomes a small algebraic problem on each flux surface, with a size determined by the number of poloidal mode numbers taken along. For each flux surface, this can be solved 
using a direct QR method. All spectra below are computed in this way, for an axial wavenumber $k=-1$ and using seven poloidal mode numbers centered around $m=1$. A larger number
of poloidal Fourier harmonics can be used, which results in a more accurate solution but also in a longer computation time. The typical radial resolution used is either 101 or 1001 
points, depending on how much detail is required in identifying each branch of the MHD continua.

\section{Cool prominence surrounded by hot medium \label{sec:cool}}
The theory presented will be applied to the two equilibrium classes presented in paper I, cool prominences embedded in hot medium and double layered prominences. 
We now discuss the first class. The equilibria and their associated profiles are discussed in our accompanying paper \citep{Blokland_2011A}. As described in our accompanying 
paper, we analyze the continuous MHD spectrum for increasing values of the gravity parameter $g$. 

\subsection{Gap creation when increasing $g$}
\begin{figure*}[ht]
  \centering
  \includegraphics[width=0.8\textwidth]{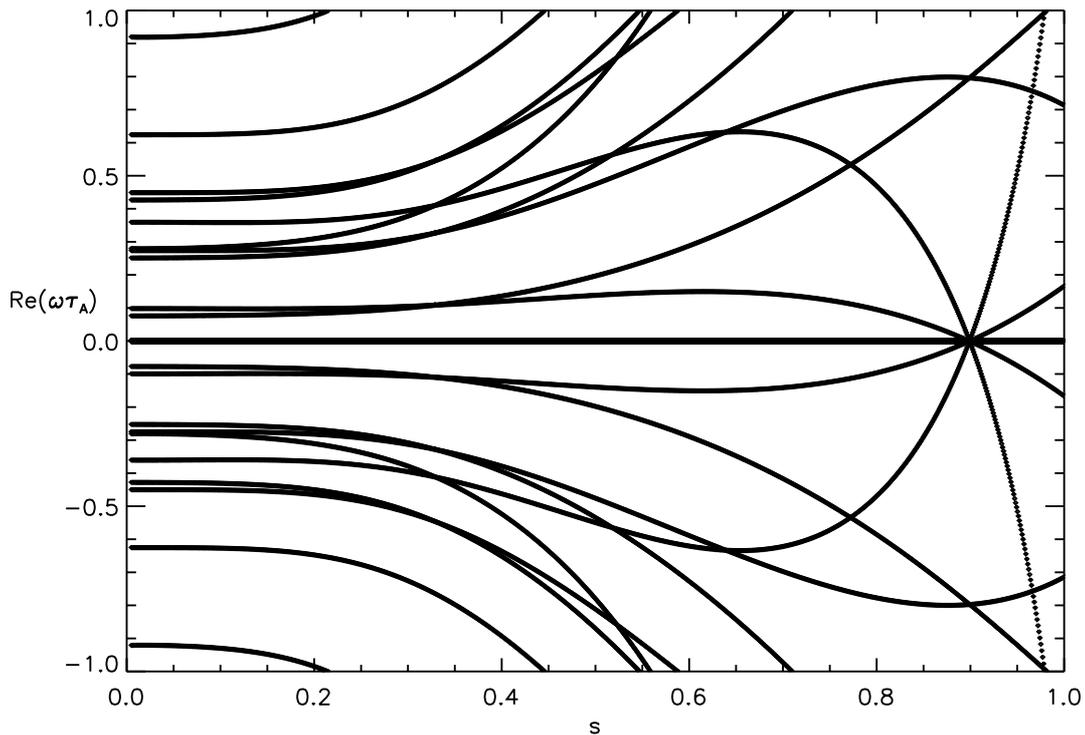}
  \caption{Part of the continuous MHD spectrum for a cool solar prominence with gravity parameter $g=0.001$ computed by PHOENIX. Shown are eigenfrequencies normalized to the
           Alfv\'en time $\tau_{\mathrm{A}}$ as a function of the radial flux coordinate $s \equiv \sqrt{\psi}$, all corresponding to flux surface localized eigenoscillations.}
  \label{fig:cool_smallg_full}
\end{figure*}

Fig.~\ref{fig:cool_smallg_full} shows part of the continuous MHD spectrum as a function of the radial flux 
coordinate $s \equiv \sqrt{\psi}$ computed by PHOENIX for a prominence equilibrium with gravity $g=0.001$ and assuming that the temperature is a flux function. The frequencies
are normalized to the Alfv\'en time~$\tau_{\mathrm{A}} \equiv L\sqrt{\mu_{0}\rho_{\mathrm{mag}}} / B_{\mathrm{mag}}$, where $\mu_{0}$, $L$, $\rho_{\mathrm{mag}}$, 
and~$B_{\mathrm{mag}}$ are the permeability, the radius of the prominence, the density on the magnetic axis, and the magnetic field on the magnetic axis, respectively. Near the 
magnetic axis where $s=0$, each branch can be labeled immediately using the analytical expressions in Eqs.~\eqref{eq:w_alfven} and \eqref{eq:w_slow} for the MHD Alfv\'en and slow 
continua. At the location $s\approx 0.9$, the safety factor $q$ of the equilibrium becomes the value of one (see paper I). At this location, the $m=1$ Alfv\'en and the $m=1$ 
slow continuum all become zero. At the low gravity value for Fig.~\ref{fig:cool_smallg_full}, no true gap can be detected, and for progressively stronger gravity, we expect 
mode coupling to occur, ultimately inducing a possible gap or even instabilities in the continua. At the same location, the $m=0$ slow and $m=2$ slow continuum have the same 
frequency, namely approximately~0.797.

\begin{figure*}[ht]
  \centering
  \includegraphics[width=\textwidth]{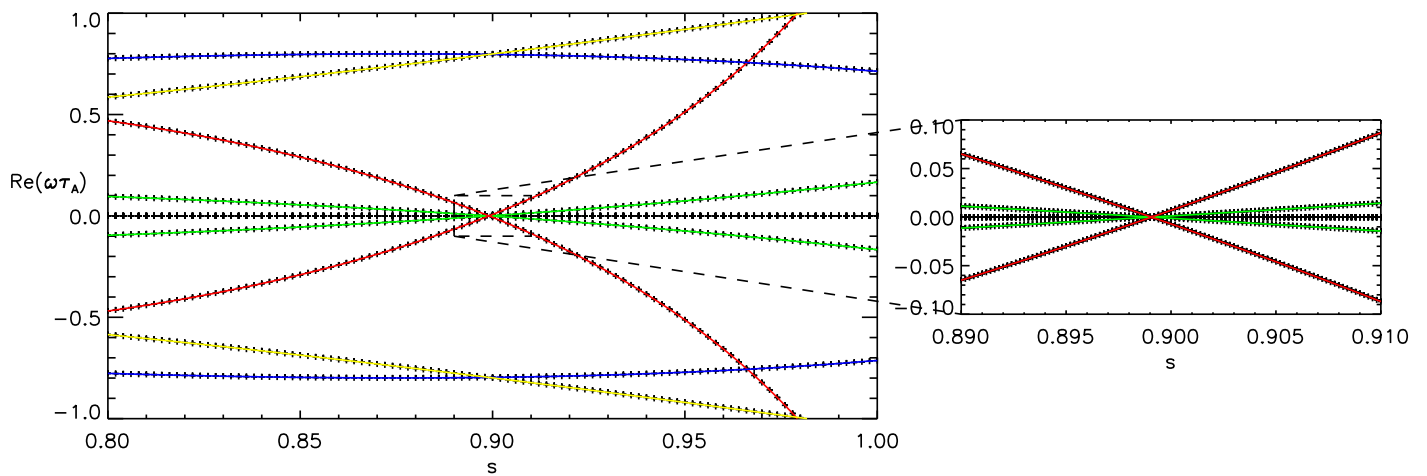}
  \caption{The real part of the sub-spectrum of the MHD continua of a cool solar prominence equilibrium with small gravity ($g=0.001$) 
           as a function of the radial flux coordinate $s \equiv \sqrt{\psi}$ for axial wavenumber $k=-1$ and poloidal mode numbers $m=\{-2,...,4\}$.
           The temperature is assumed to be a flux function.
           The crosses are PHOENIX results, while the colored lines compare them with the four-mode coupling scheme. The colors red, green, blue, and yellow
           correspond to dominant $\overline{\eta}_{1}$, $\overline{\zeta}_{1}$, $\overline{\zeta}_{0}$, and $\overline{\zeta}_{2}$ components of the eigenvector, respectively.}
  \label{fig:cool_smallg}
\end{figure*}

We next take a closer look at the continuous MHD spectrum about the $q=1$ surface, to compare the PHOENIX results with the ones predicted by the four-mode coupling 
scheme in Eq.~\eqref{eq:smallg_deltam}. This comparison is presented in Fig.~\ref{fig:cool_smallg}. Results from PHOENIX are represented by crosses, while the colored lines 
result from solving the eigenvalue equation in Eq.~\eqref{eq:smallg_deltam}. The colors indicate the dominant characters, which are easily determined from the associated eigenvector. 
The colors red, green, blue, and yellow correspond to dominant $\overline{\eta}_{1}$ (Alfv\'en), $\overline{\zeta}_{1}$ (slow), $\overline{\zeta}_{0}$ (slow), 
and $\overline{\zeta}_{2}$ (slow) characters, respectively. The plot shows that there is excellent agreement between both methods, as expected for small gravity. Furthermore, 
the $m=1$ Alfv\'en continuum (red) intersects the $m=0$ (blue) and $m=2$ slow continuum (yellow). This intersection may lead to mode coupling, resulting in a $\Delta m=1$ gap. 
The analysis of this kind of gap is left to future work. The crosses at zero frequency correspond to the Eulerian entropy continuum, as also found numerically by PHOENIX.

\begin{figure*}[ht]
  \centering
  \includegraphics[width=\textwidth]{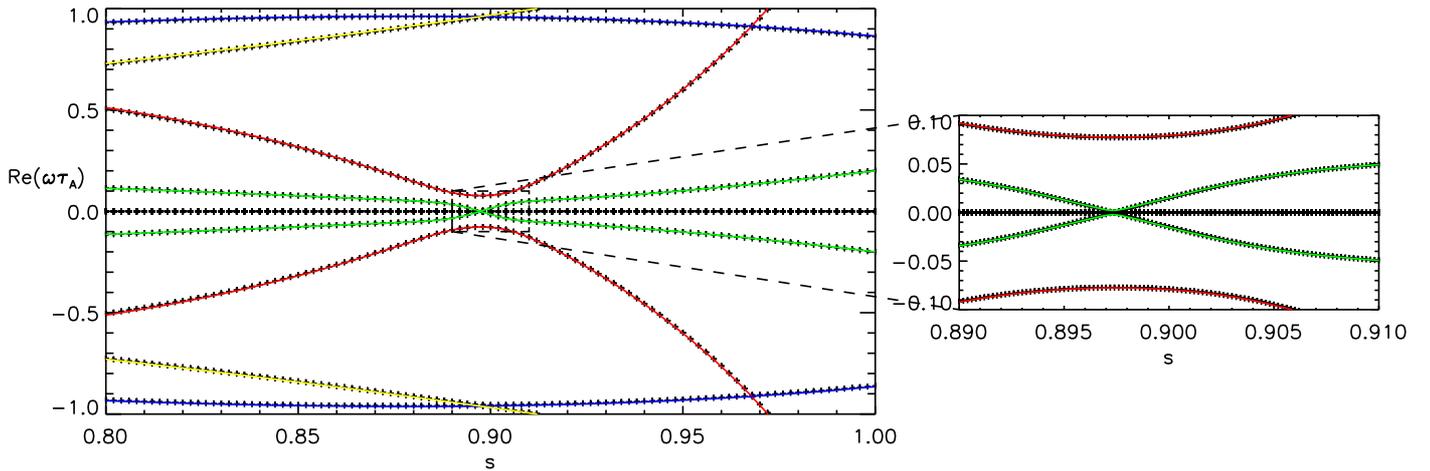}
  \caption{The real part of the sub-spectrum of the MHD continua of a cool solar prominence equilibrium with intermediate gravity $g=0.500$ 
           as a function of the radial flux coordinate $s \equiv \sqrt{\psi}$ for axial wavenumber $k=-1$ and poloidal mode numbers $m=\{-2,...,4\}$.
           Temperature is assumed to be a flux function.
           Crosses are the PHOENIX results, while the colored lines give the four-mode coupling scheme. The colors red, green, blue, and yellow
           correspond to the dominant $\overline{\eta}_{1}$, $\overline{\zeta}_{1}$, $\overline{\zeta}_{0}$, and $\overline{\zeta}_{2}$ components of the eigenvector.}
  \label{fig:cool_intermediateg}
\end{figure*}
\begin{figure*}[ht]
  \centering
  \includegraphics[width=\textwidth]{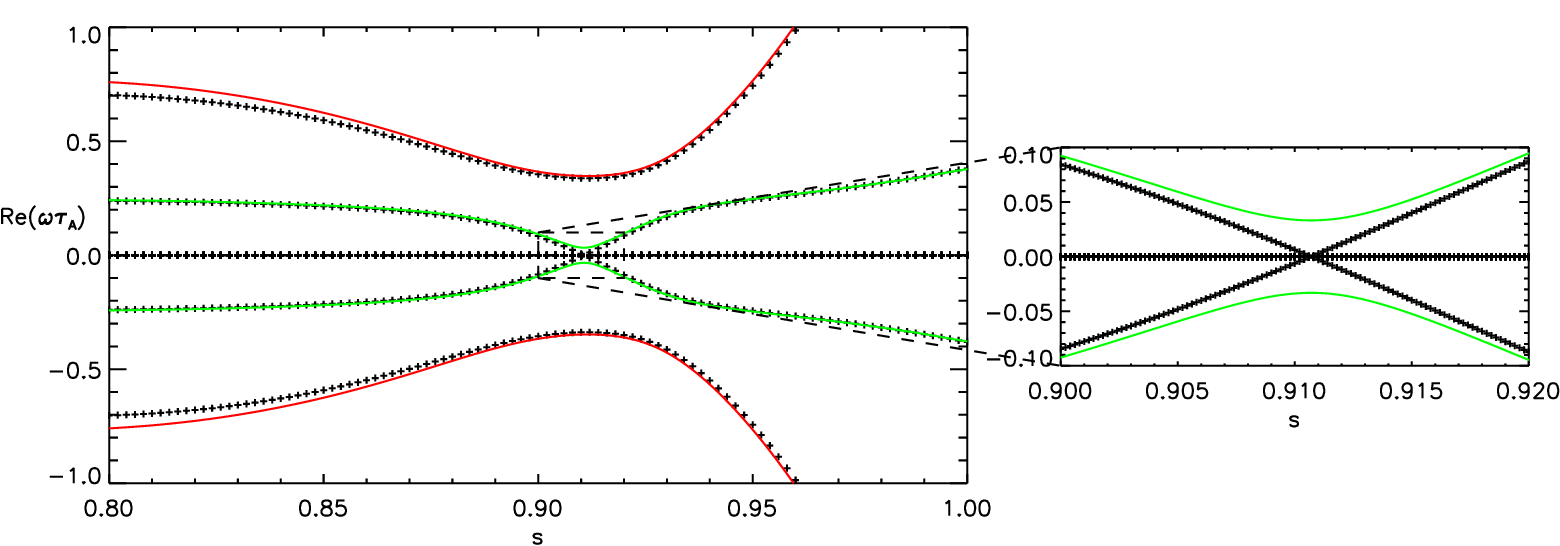}
  \caption{The real part of the sub-spectrum of the MHD continua of a cool solar prominence equilibrium with strong gravity ($g=1.000$)
           as a function of the radial flux coordinate $s \equiv \sqrt{\psi}$ are shown for axial wavenumber $k=-1$ and poloidal mode numbers $m=\{-2,...,4\}$.
           Here, the temperature is assumed to be a flux function.
           The crosses are the PHOENIX results, while the colored lines are the results of the four-mode coupling scheme. The color red, green, blue, and yellow
           correspond to the dominant $\overline{\eta}_{1}$, $\overline{\zeta}_{1}$, $\overline{\zeta}_{0}$, and $\overline{\zeta}_{2}$ components of the eigenvector.}
  \label{fig:cool_strongg}
\end{figure*}

We now repeat the same analysis, for a case with intermediate gravity $g=0.500$ and still assuming that the temperature is a flux function. The results of PHOENIX results are
in excellent agreement with the solutions of the four-mode coupling scheme in Eq,~\eqref{eq:smallg_deltam}, as shown in Fig.~\ref{fig:cool_intermediateg}. 
For this larger $g$ value, the results now clearly illustrate that a gap appears in the $m=1$ Alfv\'en continuum, while no gap appears in the $m=1$ slow continuum. We note that 
the radial flux coordinate $s$ of the $q=1$ surface is shifted inwards relative to the previously discussed equilibrium. However, the radial location $r$ of this surface remains 
almost the same (see paper I).

If we continue to increase the gravity parameter $g$, for a cool prominence equilibrium where temperature is a flux function, we find progressive deviations between numerical 
and (approximate) analytical predictions. The plot presented in Fig.~\ref{fig:cool_strongg} shows the results from PHOENIX (crosses) and the four-mode coupling scheme (colored lines).
Near $q=1$, the results of the four-mode coupling scheme corresponding to the Alfv\'en continuum (red) show good agreement with PHOENIX, while for a slow continuum (green) there is 
a discrepancy. The four-mode coupling scheme indicates that a gap in the slow continuum may result, while PHOENIX does not find such a gap. The gap in the Alfv\'en continuum 
clearly becomes larger if gravity is increased. The radial location $r$ of the $q=1$ surface remains almost the same as in the two previously discussed cases, while its radial 
flux coordinate $s$ is shifted outward.

\begin{figure*}[ht]
  \centering
  \includegraphics[width=0.7\textwidth]{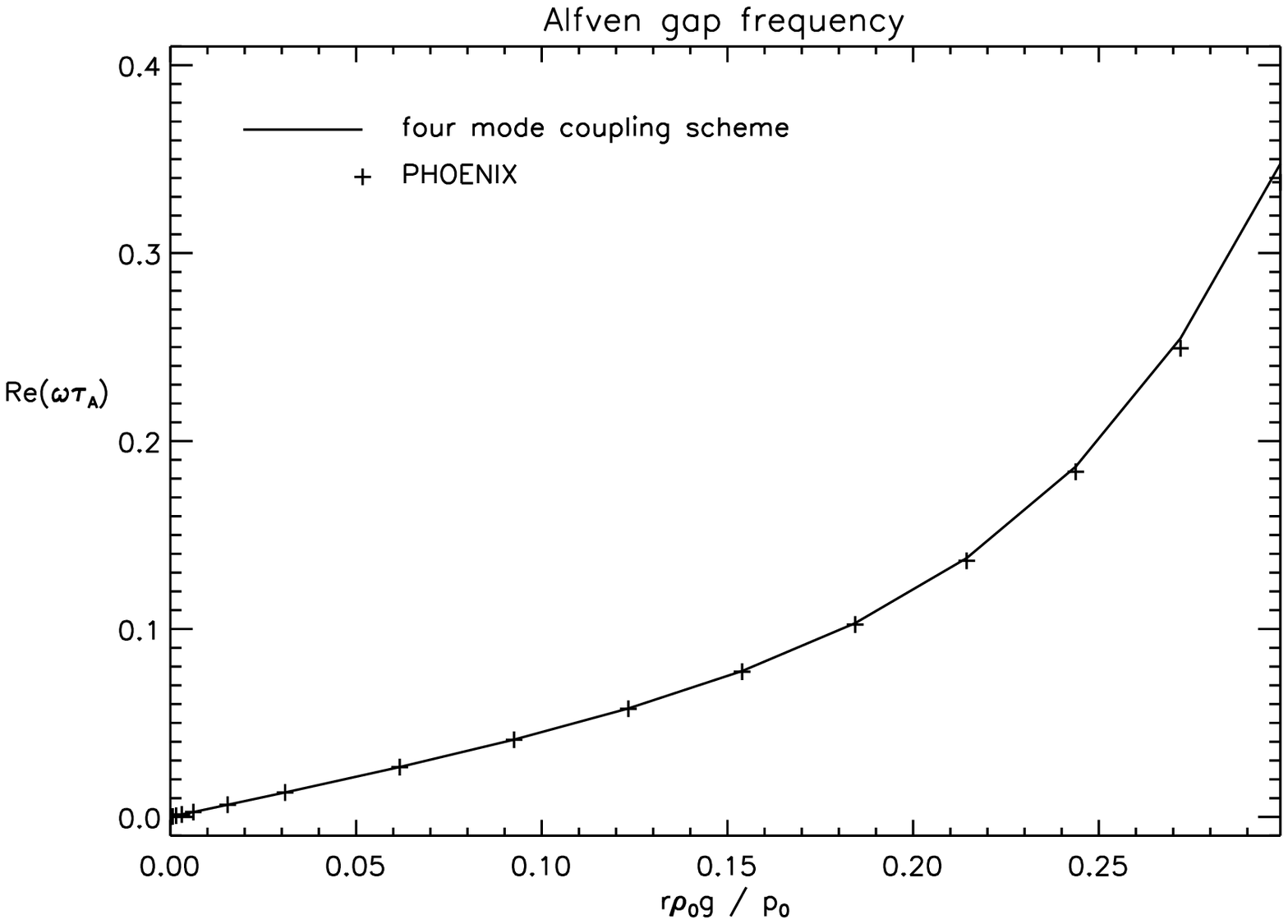}
  \caption{The frequency of the Alfv\'en continuum mode $\overline{\eta}_{1}$ at the resonant $q=1$ surface as a function of the gravity. The temperature is assumed to
           be a flux function.}
  \label{fig:cool_Alfvengap}
\end{figure*}

In the previous three equilibria, which only differ in terms of the increased role of gravity, we observed that the gap appearing in the Alfv\'en continuum becomes larger 
if the gravity is increased. This dependence was investigated by determining the Alfv\'en frequency at the $q=1$ surface, for various values of the gravity parameter. 
The result of this parametric investigation (involving the construction of numerical equilibria, subsequently diagnosed for their MHD continua by PHOENIX) has been plotted 
in Fig.~\ref{fig:cool_Alfvengap}. The plot shows excellent agreement between PHOENIX and the growth of the `gap' predicted by the four-mode coupling scheme. It is seen that the 
Alfv\'en gap frequency depends non-linearly on the gravity parameter $g$. Despite this excellent agreement for the Alfv\'en continuum gap up to $g=1$, successive 
differences occur, e.g. as noted for the slow continua shown in Fig.~\ref{fig:cool_strongg}. In Fig.~\ref{fig:cool_Alfvengap}, the horizontal axis does not use the gravity 
parameter $g$ itself, but rather a scaled value. The equilibrium with gravity $g=0.001$, $g=0.500$, and $g=1.000$ correspond to $r\rho_{0}g/p_{0}=0.0003$, $r\rho_{0}g/p_{0}=0.154$,
and $r\rho_{0}g/p_{0}=0.300$, respectively.

\subsection{Unstable continua}
\begin{figure*}[ht]
  \centering
  \includegraphics[width=\textwidth]{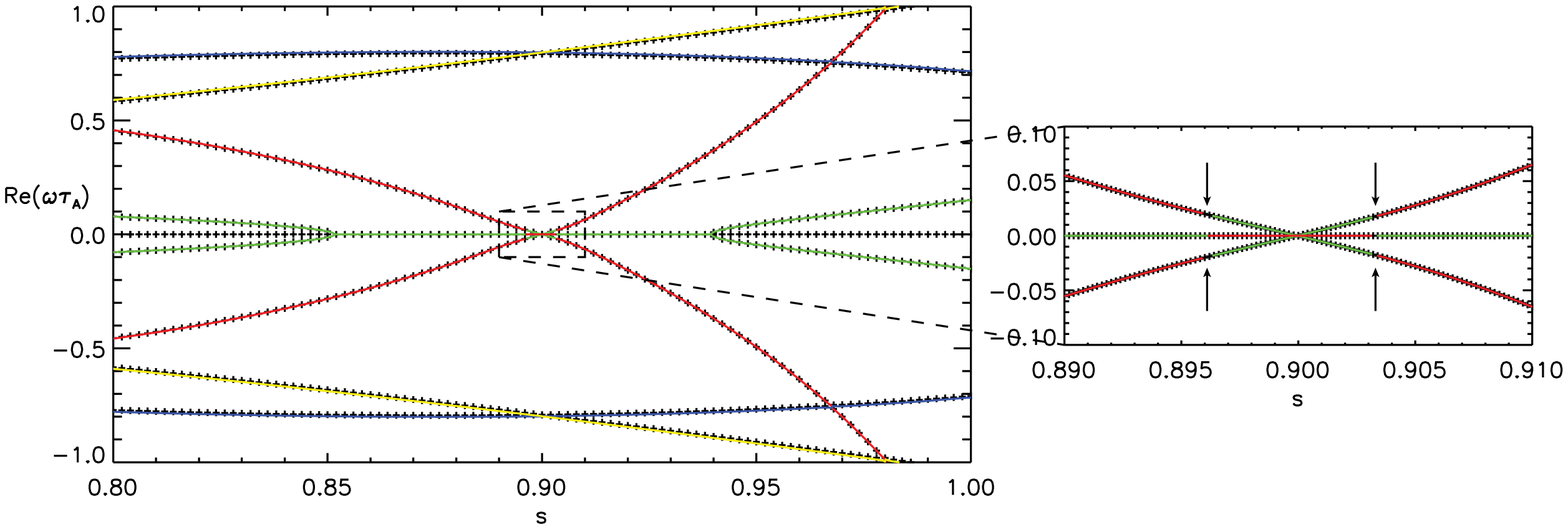}
  \caption{The real part of the sub-spectrum of the MHD continua of a cool solar prominence equilibrium with intermediate gravity ($g=0.500$)
           as a function of the radial flux coordinate $s \equiv \sqrt{\psi}$ are shown for axial wavenumber $k=-1$ and poloidal mode numbers $m=\{-2,...,4\}$.
           Density is assumed to be a flux function.
           The crosses are the PHOENIX results, while the colored lines are the results of the four-mode coupling scheme. The color red, green, blue, and yellow
           correspond to the dominant $\overline{\eta}_{1}$, $\overline{\zeta}_{1}$, $\overline{\zeta}_{0}$, and $\overline{\zeta}_{2}$ component of the eigenvector.}
  \label{fig:cool_density_Re}
\end{figure*}
\begin{figure*}[ht]
  \centering
  \includegraphics[width=\textwidth]{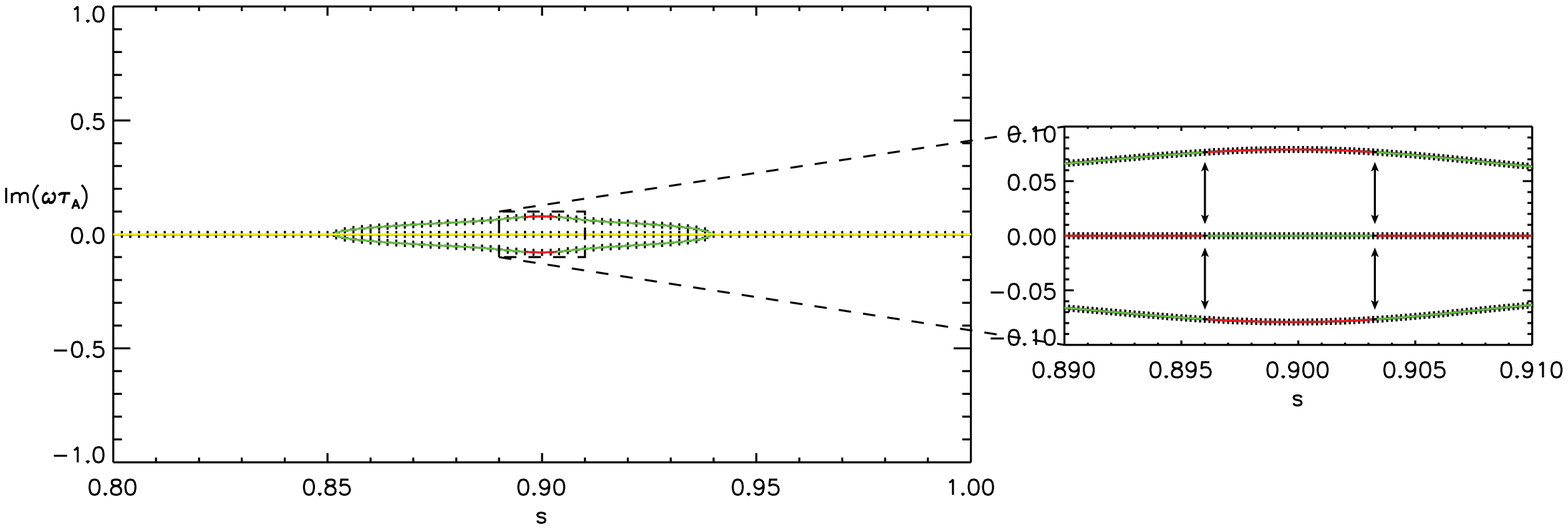}
  \caption{The imaginary part of the sub-spectrum of the MHD continua of a cool solar prominence equilibrium with intermediate gravity ($g=0.500$)
           as a function of the radial flux coordinate $s \equiv \sqrt{\psi}$ are shown for axial wavenumber $k=-1$ and poloidal mode numbers $m=\{-2,...,4\}$. 
           These quantify the growth rates of the CCI instabilities.
           Density is assumed to be a flux function.
           The crosses are the PHOENIX results, while the colored lines are the results of the four-mode coupling scheme. The color red, green, blue, and yellow
           correspond to the dominant $\overline{\eta}_{1}$, $\overline{\zeta}_{1}$, $\overline{\zeta}_{0}$, and $\overline{\zeta}_{2}$ component of the eigenvector.}
  \label{fig:cool_density_Im}
\end{figure*}
Finally, to illustrate the effect of the chosen flux function in the equilibrium on the MHD continua, we now discuss in the intermediate gravity case $g=0.500$, the continuous spectra 
when the density or the entropy is a flux function. We recall that when the density is a flux function the continuous MHD spectrum may become unstable. The continuous MHD spectrum 
of the density case is presented in Fig.~\ref{fig:cool_density_Re}. The figure shows there is again an excellent agreement between the four-mode coupling scheme and PHOENIX. The 
plot shows two distinct features, that are not present when the temperature is a flux function. Firstly, no gap appears in the $m=1$ Alfv\'en continuum. Near the $q=1$  surface 
($s \approx 0.90$), an interesting observation can be made, where Alfv\'en and slow mode characters mingle. A mode at for example $s=0.89$ with dominant $m=1$ Alfv\'en character 
slowly changes to a dominant $m=1$ slow character when it is traced closer to the $q=1$ surface. The transition points are indicated by the arrows in the plot. The second distinct 
feature is the merger of the two $m=1$ slow branches to one at $s \approx 0.852$ and $s \approx 0.94$. Exactly at these locations, the $m=1$ slow continuum turns unstable as shown in 
Fig.~\ref{fig:cool_density_Im}. This figure quantifies the imaginary parts of the continuum eigenfrequencies, corresponding to (normalized) growth rates. This instability is the 
convective continuum instability (CCI) as discussed by \cite{Blokland_2007B}. There is again excellent agreement between PHOENIX and the four-mode coupling scheme. Furthermore, 
the transition points of the dominant character of the mode indicated by arrows in Fig.~\ref{fig:cool_density_Im}  are exactly at the same locations as those indicated in 
Fig.~\ref{fig:cool_density_Re}. From this analysis, we conclude that the CCI at the $q=1$ surface has a dominant Alfv\'en character.

\begin{figure*}[ht]
  \centering
  \includegraphics[width=0.7\textwidth]{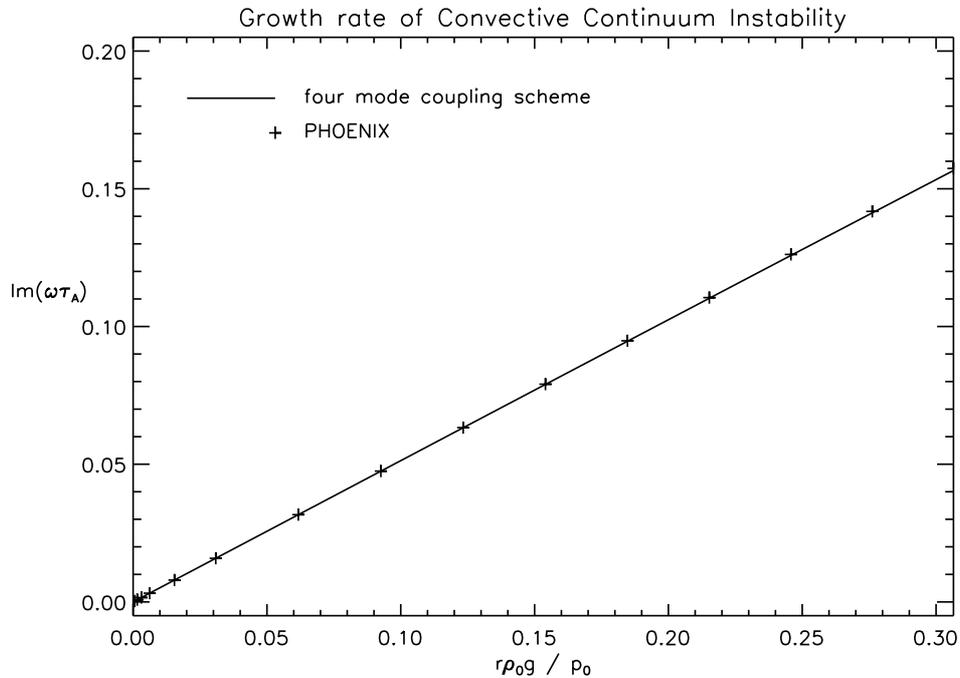}
  \caption{The growth rate of the CCI mode at the resonant $q=1$ surface as a function of the gravity parameter. The density is assumed to be a flux function.}
  \label{fig:cool_CCI}
\end{figure*}

As in the case where temperature is a flux function, we can parametrically investigate the dependence of the growth rate of the CCI on the gravity parameter $g$. This analysis again 
involves repeated FINESSE computations for increasing values of $g$, all diagnosed with PHOENIX, and the extraction of the growth rate for the most unstable mode in each case. 
These results are presented in Fig.~\ref{fig:cool_CCI}. Once more, there is excellent agreement between the two methods. Furthermore, our analysis clearly identities a linear 
dependence between the growth rate and the gravity parameter $g$. Owing to our non-dimensionalization, this implies that there is a larger growth of the CCI mode for 
flux ropes with embedded prominences that (1) gradually grow in cross-sectional extent, and/or (2) become more heavy (density increase), and/or (3) decrease in magnetic field strength.

In the context of prominence dynamics, the existence of equilibria viable to unstable continua may have relevance to the complex dynamics observed in high resolution 
prominence movies~\citep{Berger08}, as well as to the sudden disappearance of prominence plasma, the so-called {\em{disparition-brusque}} 
phenomenon~\citep[see e.g.][and references therein]{tandberg2011}. It should be emphasized that the onset of the CCI instability will be localized on those flux surfaces where 
the CCI criterion is violated. As shown in Fig.~\ref{fig:cool_density_Im}, growth rates can vary from one flux surface to another in a given unstable case, while in addition, 
the  maximal growth rate also varies because of gradual equilibrium changes, as demonstrated in Fig.~\ref{fig:cool_CCI}. Whether this linear prediction for the onset of internal, 
continuum-driven dynamics leads to the rather turbulent prominence dynamics revealed by contemporary observations, or even to full disruption and the disappearance of prominence 
structures, must be answered by follow-up high resolution, nonlinear numerical simulations. This requires initializing multi-dimensional nonlinear simulations with the spectrally 
diagnosed equilibria.

\subsection{Spectra for equilibria in which entropy is a flux function}
\begin{figure*}[ht]
  \centering
  \includegraphics[width=\textwidth]{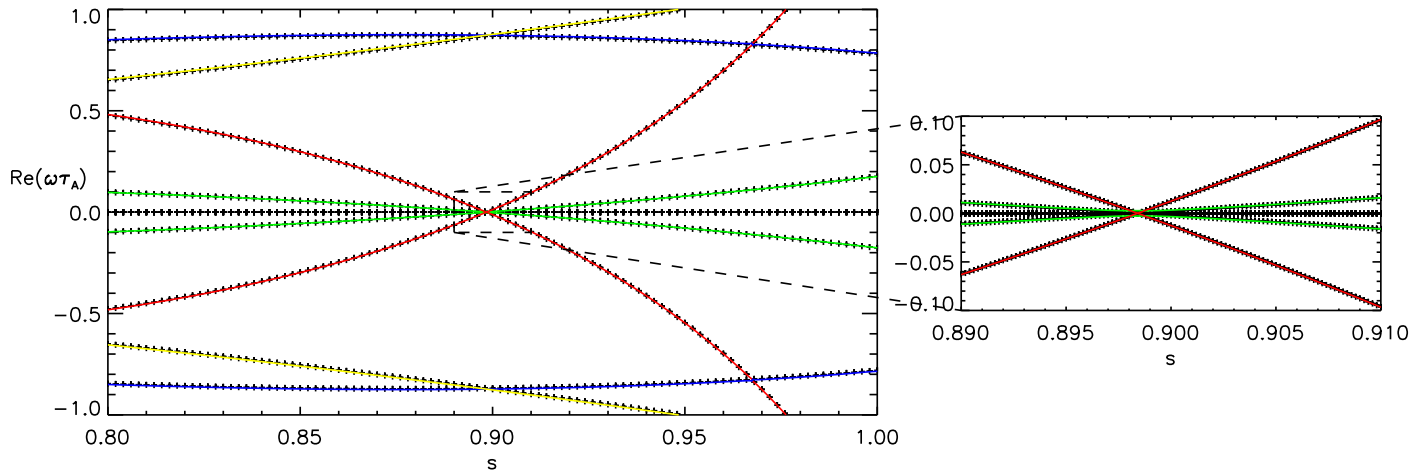}
  \caption{The real part of the sub-spectrum of the MHD continua of a cool solar prominence equilibrium with intermediate gravity ($g=0.500$)
           as a function of the radial flux coordinate $s \equiv \sqrt{\psi}$ are shown for axial wavenumber $k=-1$ and poloidal mode numbers $m=\{-2,...,4\}$.
           Here, entropy is assumed to be a flux function.
           The crosses are the PHOENIX results, while the colored lines are the results of the four-mode coupling scheme. The color red, green, blue, and yellow
           correspond to the dominant $\overline{\eta}_{1}$, $\overline{\zeta}_{1}$, $\overline{\zeta}_{0}$, and $\overline{\zeta}_{2}$ component of the eigenvector.}
  \label{fig:cool_entropy}
\end{figure*}

We now consider the case where entropy is a flux function, for an equilibrium with intermediate gravity ($g=0.500$). The resulting continuous MHD spectrum is presented in 
Fig.~\ref{fig:cool_entropy}. There is again excellent agreement between both methods. It can be clearly seen that no gap in the Alfv\'en continuum appears. The continua for 
the strong gravity $g=1.000$ case have also been computed, which also show no appearance of a gap. This can be understood by examining Eq.~\eqref{eq:polynomial} for the 
eigenfrequency. When the entropy is a flux function, the Brunt-V\"ais\"al\"a frequency projected on a flux surface $N_{\mathrm{BV,pol}}$ is zero, thereby transforming the 
eigenfrequency given by Eq.~\eqref{eq:polynomial}, to one similar to a case without gravity, for which no gap can appear.

\section{Double layered prominences \label{sec:double}}
We discuss the continuous MHD spectrum of the second equilibrium class, the double layered prominences. As for cool prominences, the equilibria
and the associated profiles are discussed in our accompanying paper \citep{Blokland_2011A}. For this equilibrium class, we also investigate the influence of the gravity on
the continuous MHD spectrum by slowly increasing the gravity. 

\subsection{Continua for increasing gravity}
\begin{figure*}[ht]
  \centering
  \includegraphics[width=0.8\textwidth]{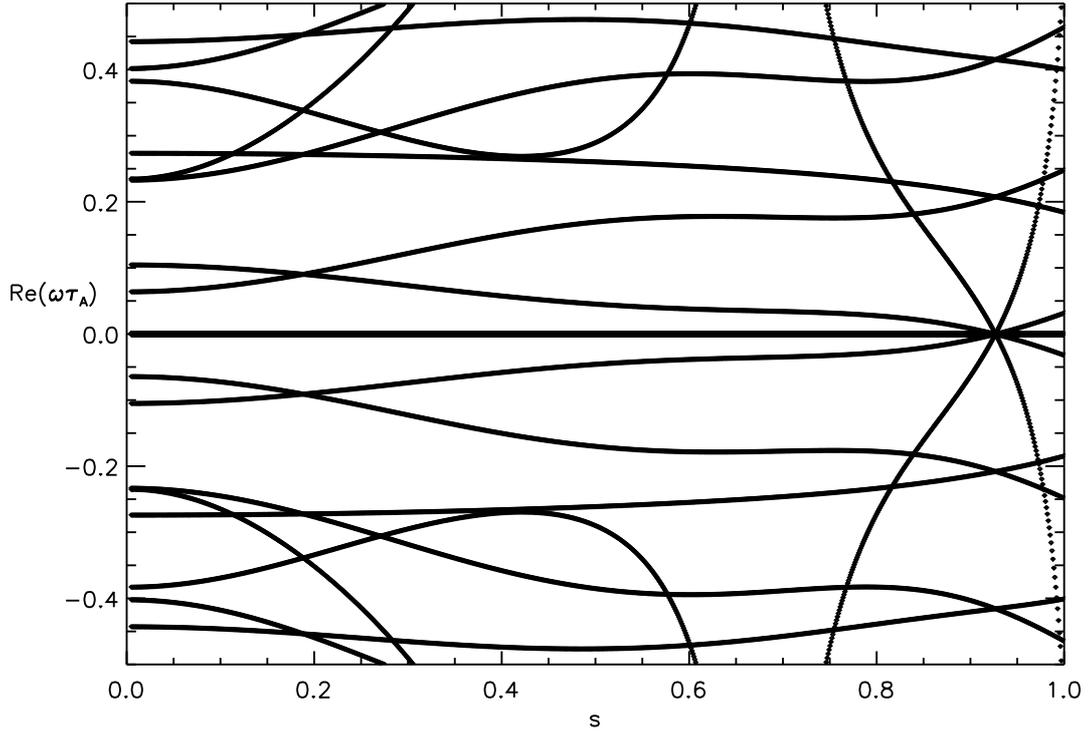}
  \caption{A part of the continuous MHD spectrum of double layered solar prominence with gravity $g=0.001$ computed by PHOENIX as a function of the radial flux 
           coordinate $s \equiv \sqrt{\psi}$.}
  \label{fig:constantT_smallg_full}
\end{figure*}

The first equilibrium that was analyzed with PHOENIX is the one where the temperature is assumed to be a flux function and the gravity is low at $g=0.001$. Part of the MHD 
spectrum is shown in Fig.~\ref{fig:constantT_smallg_full}. The individual branches near the magnetic axis can be labeled using the analytical expressions for the Alfv\'en 
continuum in Eq.~\eqref{eq:w_alfven} and slow continuum in Eq.~\eqref{eq:w_slow}. The associated radial flux coordinate $s$ of the $q=1$ surface is approximately 0.927. At exactly
this location, the $m=1$ Alfv\'en and $m=1$ slow continuum are zero, as expected, and the $m=0$ and $m=2$ slow continuum have the same frequency.

\begin{figure*}[ht]
  \centering
  \includegraphics[width=\textwidth]{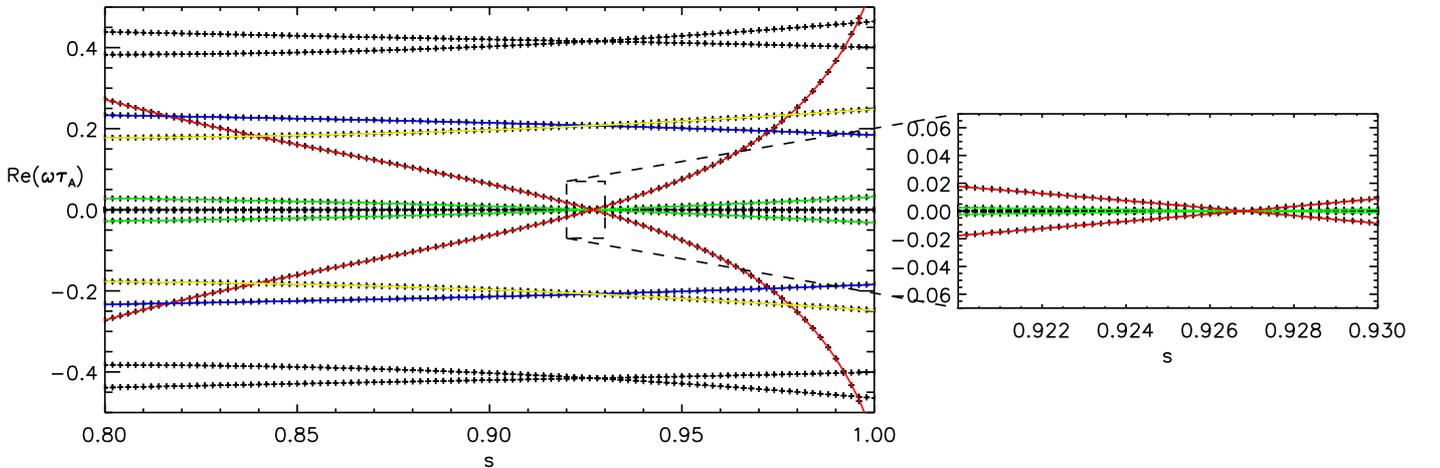}
  \caption{The real part of the sub-spectrum of the MHD continua of a double layered solar prominence equilibrium with small gravity ($g=0.001$)
           as a function of the radial flux coordinate $s \equiv \sqrt{\psi}$ are shown for axial wavenumber $k=-1$ and poloidal mode numbers $m=\{-2,...,4\}$.
           Here, the temperature is assumed to be a flux function.
           The crosses are the PHOENIX results, while the colored lines are the results of the four-mode coupling scheme. The color red, green, blue, and yellow
           correspond to the dominant $\overline{\eta}_{1}$, $\overline{\zeta}_{1}$, $\overline{\zeta}_{0}$, and $\overline{\zeta}_{2}$ component of the eigenvector.}
  \label{fig:constantT_smallg}
\end{figure*}
The plot shown in Fig.~\ref{fig:constantT_smallg} contains the PHOENIX results (crosses) together with the solutions of the four-mode coupling scheme represented by
Eq.~\eqref{eq:smallg_deltam} (colored lines), and illustrates the excellent agreement between the two methods. As for the cool prominences, the $m=1$ Alfv\'en continuum (red) 
intersects the $m=0$ (blue) and $m=2$ slow continuum (yellow). This may in turn lead to mode coupling, creating a high frequency $\Delta m=1$ gap. The analysis of these 
$\Delta m = 1$ gaps was not performed here, but left to future work. We note that the crosses with zero frequency correspond to the Eulerian entropy continuum.

\begin{figure*}[ht]
  \centering
  \includegraphics[width=\textwidth]{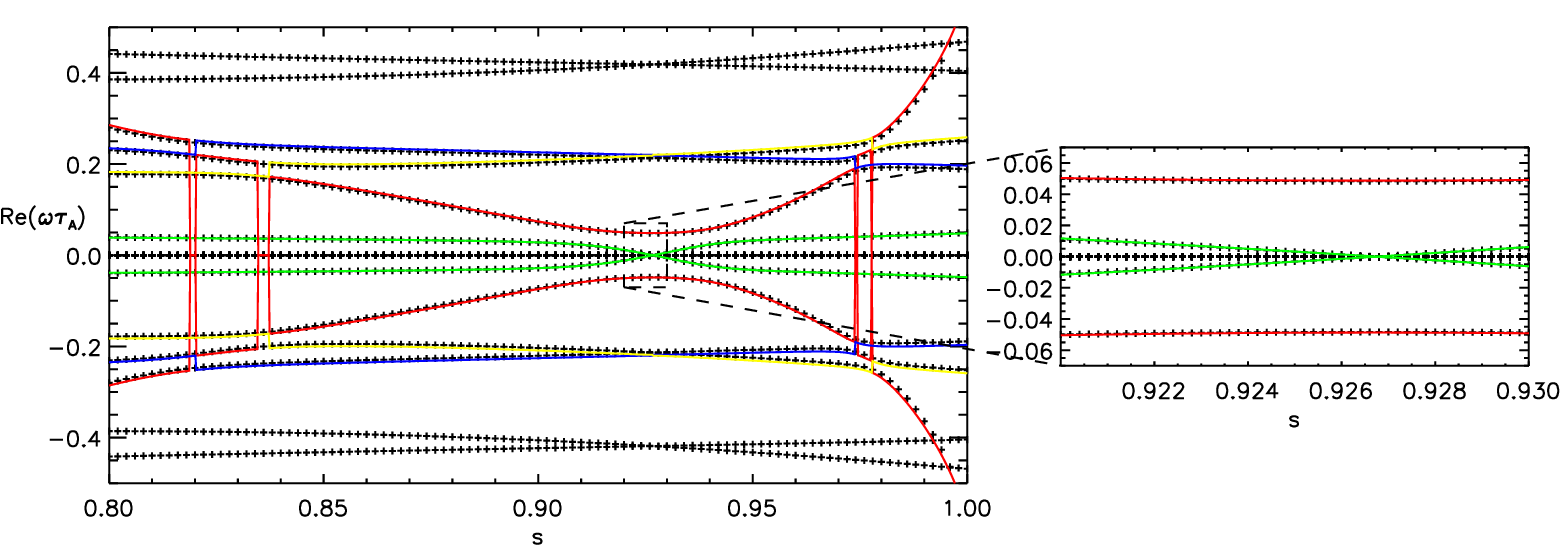}
  \caption{The real part of the sub-spectrum of the MHD continua of a double layered solar prominence equilibrium with strong gravity ($g=0.500$)
           as a function of the radial flux coordinate $s \equiv \sqrt{\psi}$ are shown for axial wavenumber $k=-1$ and poloidal mode numbers $m=\{-2,...,4\}$.
           Here, the temperature is assumed to be a flux function.
           The crosses are the PHOENIX results, while the colored lines are the results of the four-mode coupling scheme. The color red, green, blue, and yellow
           correspond to the dominant $\overline{\eta}_{1}$, $\overline{\zeta}_{1}$, $\overline{\zeta}_{0}$, and $\overline{\zeta}_{2}$ component of the eigenvector.}
  \label{fig:constantT_intermediateg}
\end{figure*}

Fig.~\ref{fig:constantT_intermediateg} contains the plot for an equilibrium with intermediate gravity $g=0.500$. The temperature is again assumed to be a flux function.
Near the $q=1$ surface ($s \approx 0.927$), the results of PHOENIX and the four-mode coupling scheme show excellent agreement for the $m=1$ Alfv\'en and slow continuum. 
Owing to the non-zero gravity, a gap has been created in the $m=1$ Alfv\'en continuum. However, $m=0$ and $m=2$ slow continua of the coupling scheme slightly 
mismatch the PHOENIX results. This mismatch is because by our neglect of the coupling with $m=-1$ and $m=3$ slow continua. In the plot, these continua
are represented by the crosses with non-zero frequency in combination without a solid line. At four different radial flux coordinates, $s \approx 0.82$, $s \approx 0.837$, 
$s \approx 0.973$, and $s\approx 0.978$ a $\Delta m=1$ gap was produced (note the color change of the solid lines about these radii). As expected, the four-mode coupling scheme 
only partly captures this kind of coupling. Therefore, about these $\Delta m=1$ gaps there is a mismatch between the results of both methods, and a separate detailed analysis 
of the $\Delta m=1$ coupling schemes appears to be required.

\begin{figure*}[ht]
  \centering
  \includegraphics[width=\textwidth]{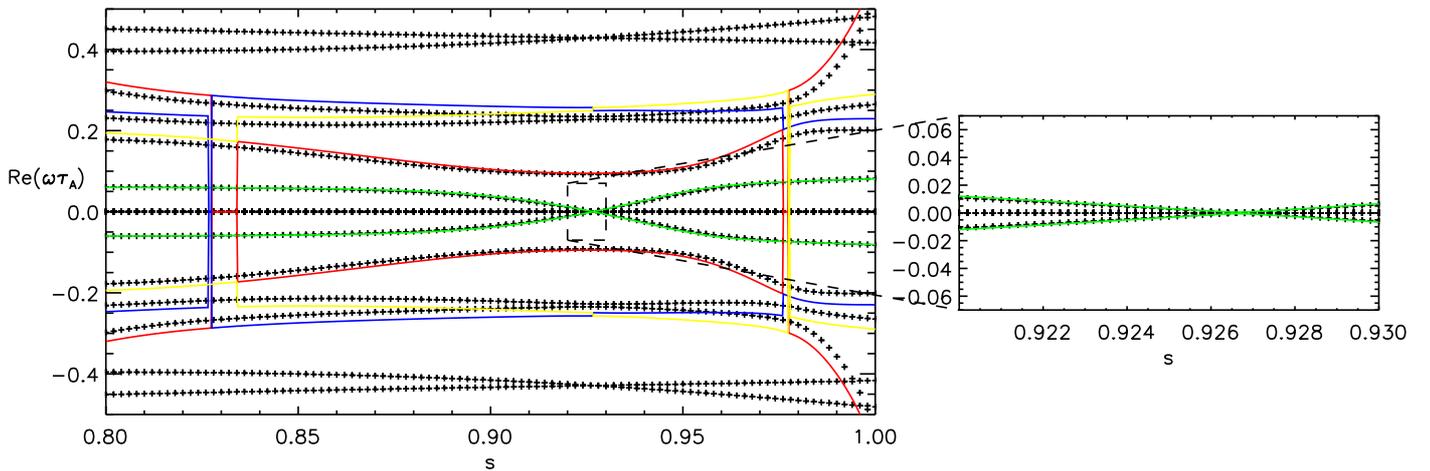}
  \caption{The real part of the sub-spectrum of the MHD continua of a double layered solar prominence equilibrium with strong gravity ($g=1.000$)
           as a function of the radial flux coordinate $s \equiv \sqrt{\psi}$ are shown for axial wavenumber $k=-1$ and poloidal mode numbers $m=\{-2,...,4\}$.
           Here, the temperature is assumed to be a flux function.
           The crosses are the PHOENIX results, while the colored lines are the results of the four-mode coupling scheme. The color red, green, blue, and yellow
           correspond to the dominant $\overline{\eta}_{1}$, $\overline{\zeta}_{1}$, $\overline{\zeta}_{0}$, and $\overline{\zeta}_{2}$ component of the eigenvector.}
  \label{fig:constantT_strongg}
\end{figure*}
The last equilibrium of this subsection that we discuss is the one with strong gravity $g=1.000$. The temperature is assumed to be a flux function. For this equilibrium, one 
observes the onset of the double layered structure in the pressure and density as demonstrated in our accompanying paper \citep{Blokland_2011A}. The results for the $m=1$
Alfv\'en and slow continuum show excellent agreement near the $q=1$ surface ($s \approx 0.927$). Owing to the stronger gravity compared to the previous case, the gap 
in $m=1$ Alfv\'en continuum is larger. Near this surface, the previously mentioned mismatch of both the $m=0$ and $m=2$ slow continuum becomes larger because of the neglect of the 
coupling with both the $m=-1$ and $m=3$ slow continuum. This coupling becomes stronger if the gravity is increased. The four-mode coupling scheme is also unable to capture the coupling 
at the radial flux coordinates $s \in [0.828, 0.833]$ and $s \in [0.975, 0.978]$. This is again because we neglected the coupling with both the $m=-1$ and $m=3$ slow continuum.
Therefore, close to these radii there is a large discrepancy between PHOENIX and the coupling scheme in Eq.~\eqref{eq:smallg_deltam}. Furthermore, we note the coupling between 
these two slow continua at the $q=1$ surface. This kind of coupling creates a $\Delta m=2$ gap that was neglected in our theoretical analysis.

\begin{figure*}[ht]
  \centering
  \includegraphics[width=0.7\textwidth]{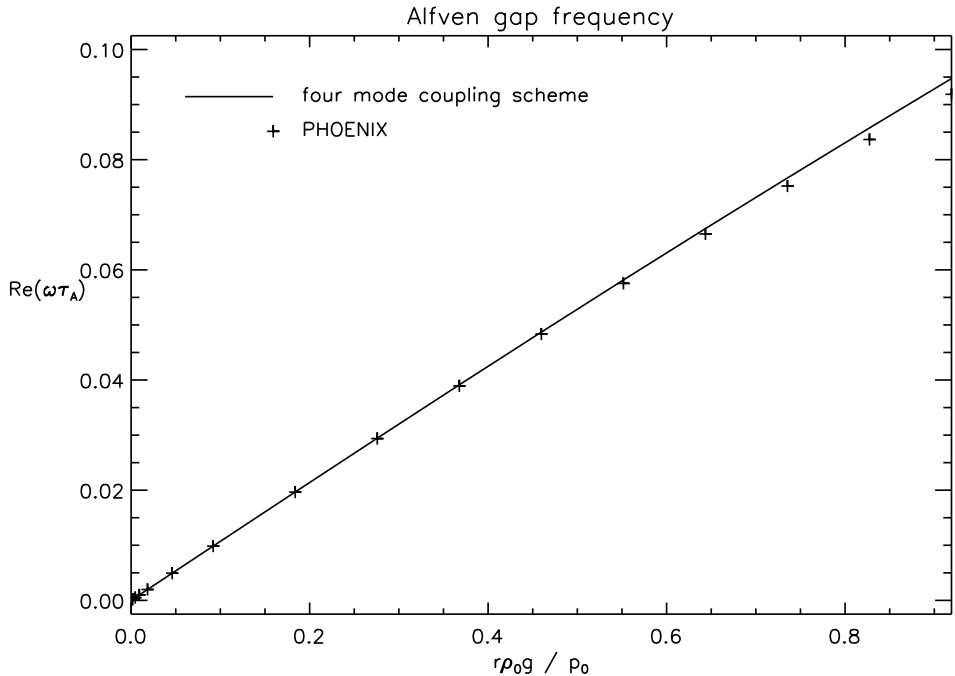}
  \caption{The frequency of the Alfv\'en continuum mode $\overline{\eta}_{1}$ at the resonant $q=1$ surface as a function of the gravity. The temperature is assumed
           to be a flux function.}
  \label{fig:constantT_Alfvengap}
\end{figure*}
As for the equilibrium class of cool prominences, we observe that when the gravity is increased, the gap in the $m=1$ Alfv\'en continuum increases. This gravity dependence of the
gap was investigated in detail by computing the Alfv\'en frequency at the $q=1$ surface for different values of the gravity. The results of this investigation are presented
in Fig.~\ref{fig:constantT_Alfvengap}. The results of PHOENIX and the four-mode coupling scheme show excellent agreement with each other. Furthermore, we note that the
frequency varies almost linearly with the gravity parameter $g$. This linear dependence is very different from the clear non-linear dependence found for cool prominences. 
This excellent agreement is again only valid for the Alfv\'en continuum. As demonstrated in Figs.~\ref{fig:constantT_intermediateg} and \ref{fig:constantT_strongg}, this does not 
automatically mean that it also holds for the slow continua. The dimensionless gravity parameter $g=0.001$, $g=0.500$, and $g=1.000$ correspond to $r\rho_{0}g/p_{0}=0.0009$, 
$r\rho_{0}g/p_{0}=0.4597$, and $r\rho_{0}g/p_{0}=0.9192$, respectively.

\subsection{Continua for strong gravity, with axial flow}
\begin{figure*}[ht]
  \centering
  \includegraphics[width=0.8\textwidth]{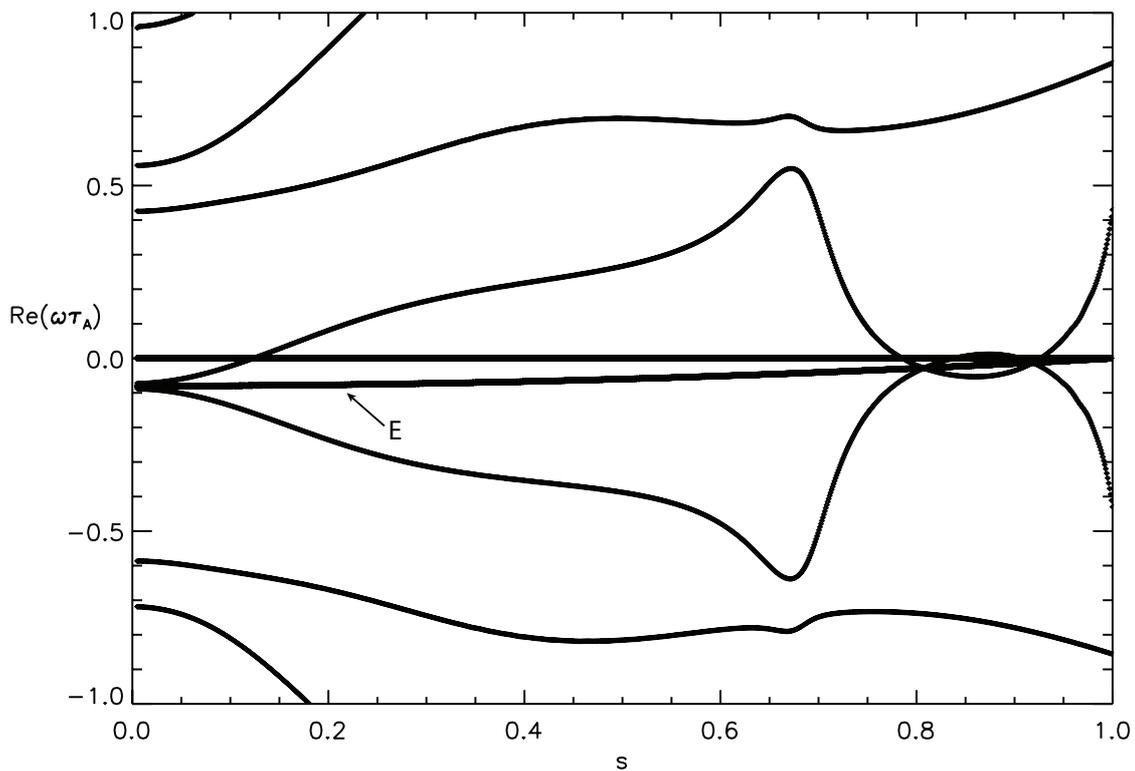}
  \caption{A part of the continuous MHD spectrum of a double layered solar prominence with strong gravity $g=5.000$ computed by PHOENIX as a function of the radial flux 
           coordinate $s \equiv \sqrt{\psi}$.}
  \label{fig:constantT_strongg_full_flow}
\end{figure*}

We now discuss the last equilibrium of the second class. For this equilibrium, we assume that the temperature is a flux function, we adopt an extra strong 
gravity parameter value ($g=5.000$), and we include an axial, shear flow. The derivation in our accompanying paper \citep{Blokland_2011A} shows that the axial flow must be a 
flux function. The flow itself is described by the profile
\begin{equation}
  v_{z}(\psi) = A_{4} ( 1 - \psi),
\end{equation}
where the constant $A_{4}=0.11$. This corresponds to a maximum Alfv\'en Mach number of 0.105. Using the typical Alfv\'en speed of~$218\, \mathrm{km}\, \mathrm{s}^{-1}$, as mentioned 
in our accompanying paper \citep{Blokland_2011A}, this translates into a filament-aligned flow of at most 22.9 km $s^{-1}$. Part of the resulting continuous MHD spectrum computed by 
PHOENIX is shown in Fig.~\ref{fig:constantT_strongg_full_flow}. Instead of using 7 poloidal harmonics, we used 15 harmonics to obtain an accurate result (i.e., one that no longer 
changed when adding even more poloidal harmonics). This is a clear indication that the strong gravity causes strong coupling between different continua. Owing the extra strong gravity,
there is no meaningful comparison between the PHOENIX results and those predicted from our theoretical analysis, which is only appropriate if the gravity is weak. In the figure, 
we indicate the non-zero Eulerian entropy continuum (where $\omega=k v_z$ labeled $E$) that breaks the up-down symmetry about the zero frequency. There is instead now symmetry 
about this Eulerian entropy continuum. Furthermore, PHOENIX finds modes with zero frequency, which correspond to the solution of the generalized eigenvalue equation in combination
with the perturbed magnetic field being divergence free. If one carefully examines the plot, one notices that one continuum branch intersects the Eulerian entropy continuum 
twice, namely at $s \approx 0.808$ and at $s \approx 0.917$. At these locations exactly, the safety factor $q$ is one. Fig.~14 of our accompanying paper \citep{Blokland_2011A} clearly 
shows the $q=1$ surface locations, as seen in a plot of safety factor $q$ across the vertical direction. Using the radial flux coordinate $s$ as a flux surface label where we go from 
the magnetic axis to the flux rope edge, there will be two $q=1$ surfaces. We note that the clear coupling effects at other radial locations as well, creating gaps that may define 
preferred frequency ranges for global eigenoscillations.

\section{Conclusions \label{sec:conclusions}}
We presented the equations describing the complete MHD spectrum. These equations have been specialized for the continuous MHD spectrum using techniques frequently
used in research on fusion plasma oscillations. From the equations for the continuous spectrum, a stability criterion was derived, showing that prominence equilibria 
for which the temperature or the entropy is a flux function have stable continuum modes. However, when the density is a flux function, the continua may turn unstable
resulting in the convective continuum instability. Whether this instability is related to the intricate dynamics witnessed in actual prominence plasmas, or explains why 
certain filaments disappear abruptly, remains to be investigated by follow-up nonlinear computations. We have demonstrated how the onset of CCI modes in flux ropes, 
which carry embedded prominences, will manifest itself as truly flux-surface localized fluctuations. Their growth rate increases when a flux rope expands (the gravity 
parameter $g$ increases because of both rope widening and the associated magnetic field weakening) or when it becomes more mass-loaded. We note that this route to potentially 
violent prominence dynamics is here related to internal prominence properties. This contrasts with studies~\citep[see e.g.][]{mactaggert09} that emphasize the role of 
external influences, studies that emphasize the response to changes in the overlying field, or the associated pressure buildup in the surroundings by the shearing of its footpoints. 
We note that the nonlinear simulations that may ultimately decide between these competing or possibly cooperating aspects, will need to go beyond the zero beta, and/or 
no gravity assumption typically employed in these studies~\citep{Aulanier02,Amari99}. Moreover, to capture the details of flux-surface localized continuum modes with 
varying MHD character (slow versus Alfv\'en) from one flux surface to another, the requirements of the 3D resolution will likely only be achievable by using effectively 
high resolution, grid-adaptive simulations. In addition, guidance on the actual magnetic field structure in prominence surroundings will be 
needed~\citep[such as that obtained in][]{schmieder07}, to distinguish between the theoretical freedom in flux profiles.

The equations for the MHD continuous spectrum have been specialized for small gravity, using an expansion in poloidal 
Fourier harmonics. The resulting equations have been applied to a low frequency $\Delta m=0$ gap near a resonant $q$ surface. Near this surface, a four-mode coupling 
scheme has been identified. This scheme couples one Alfv\'en mode and three slow modes, and can cause gaps (avoided crossings) between continuum branches as these are 
traced across all nested flux surfaces. These gaps appear at low frequency, and form the ideal locations for global discrete eigenoscillations. It remains to be shown 
which kind of coupling occurs at finite frequency, $\Delta m=1$ gaps, and to determine whether discrete global modes exist within these frequency ranges, under 
realistic prominence conditions. These global modes, as well as their coupling to flux localized surface modes (with resonant damping), need further study, to address 
whether previously established prominence seismology results are consistent with these more realistic prominences.

The MHD spectral code PHOENIX has been used to accurately compute the continuous MHD spectrum of prominence equilibria. Two classes of equilibria have been considered,
cool prominences embedded in hot medium and prominences with a double layered pressure and density structure for different values of the dimensionless gravity 
parameter as discussed in our accompanying paper \citep{Blokland_2011A}. The results of PHOENIX are compared with analytic results of the four-mode coupling scheme, and we have found 
excellent agreement for small gravity (as expected), while for strong gravity, we have identified progressive discrepancies, emphasizing the need for full numerical diagnosis. For cool 
prominences, this discrepancy manifested itself in the slow continuum with the same poloidal mode number as the Alfv\'en continuum, while for double layered prominences, differences 
occurred in the slow continuum with a poloidal mode number $\pm$1 with respect to the Alfv\'enic one. For both classes, when temperature is a flux function, the results demonstrate that 
a gap in the Alfv\'en continuum can appear due to the gravity. For cool prominences, the Alfv\'en frequency associated with the gap depends non-linearly on the gravity parameter, while 
for double layered prominences it scales linearly with it.

For cool prominences for which the density or the entropy is a flux function, the continuous MHD spectrum has also been computed. In the case of the density, the continuous
spectrum becomes unstable leading to convective continuum instability. Our analysis reveals that close to the $q=1$ surface, the CCI mode has a dominant Alfv\'en character, 
which changes away from this surface to a dominant slow type. Furthermore, no gap in the Alfv\'en continuum appears, in contrast to the case when the temperature is a flux function.
For the entropy case, the continuous spectrum is stable and again no gap appears in the Alfv\'en continuum. In the context of prominence configurations, long-lived equilibria almost 
necessitate the temperature flux function dependence, but for active region prominences, the findings about CCI possibilities may provide important information about the onset
of turbulent dynamics.

\begin{acknowledgements}
This work was carried out within the framework of the European Fusion Programme, supported by the European Communities under contract of the Association EURATOM/FOM. 
Views and opinions expressed herein do not necessarily reflect those of the European Commission. 
RK acknowledges financial support by project GOA/2009/009 (K.U.Leuven). The research leading to these results has received 
funding from the European Commission's Seventh Framework Programme (FP7/2007-2013) under the grant agreement SWIFF (project nr 263340, www.swiff.eu).
\end{acknowledgements}

\bibliographystyle{aa}
\bibliography{references}

\end{document}